\documentclass[a4paper,fleqn,usenatbib,onecolumn,useAMS]{mnras}

\newif\ifAMStwofonts

\usepackage{graphicx}	
\usepackage{amsmath}	
\usepackage{amssymb}	
\usepackage{epsfig}
\usepackage{color}

\def\pmb#1{\mbox{\boldmath$#1$}}
\def\gtsim {>\kern-1.2em\lower1.1ex\hbox{$\sim$}}
\def\ltsim {<\kern-1.2em\lower1.1ex\hbox{$\sim$}}
\def\gtsim {>\kern-1.2em\lower1.1ex\hbox{$\sim$}}
\def\ltsim {<\kern-1.2em\lower1.1ex\hbox{$\sim$}}

\def\be{\begin{equation}}
\def\ee{\end{equation}}

\def\rmi{{\rm i}}

\begin{document}

\title[Convective Core Excites Non-Radial Pulsations]{Rotating Convective Core Excites Non-Radial Pulsations
to Cause Rotational Modulations in Early-Type Stars}

\author[U. Lee, H. Saio]{
Umin Lee$^{1}$\thanks{E-mail: lee@astr.tohoku.ac.jp} and Hideyuki Saio$^{1}$\thanks{E-mail: saio@astr.tohoku.ac.jp}
\\
$^{1}$Astronomical Institute, Tohoku University, Sendai, Miyagi 980-8578, Japan\\
}

\date{Accepted XXX. Received YYY; in original form ZZZ}
\pubyear{2015}

\maketitle

\begin{abstract}
We discuss low-frequency g modes excited by resonant couplings with weakly unstable oscillatory convective modes in the rotating convective core in early-type main-sequence stars. Our non-adiabatic pulsation analyses including the effect of Coriolis force for $2\,M_\odot$ main-sequence models show that if the convective core rotates slightly faster than the surrounding radiative layers, g modes in the radiative envelope are excited by a resonance coupling. The frequency of the excited g mode in the inertial frame is close to $|m\Omega_{\rm c}|$ with $m$ and $\Omega_{\rm c}$ being the azimuthal order of the g mode and the rotation frequency of the convective core, respectively. These g mode frequencies are consistent with those of photometric rotational modulations and harmonics observed in many early-type main-sequence stars.
In other words, these g-modes provide a non-magnetic explanation for the rotational light modulations detected in many early type main-sequence stars.

\end{abstract}



\section{introduction}

Every early-type main-sequence star has a convective core, in which a small perturbation grows monotonically in the absence of rotation (sometimes called an unstable $g^{-}$ mode (e.g., \citealt{Cox1980})). In the presence of rotation, however, the perturbation oscillates with amplitude growing exponentially (e.g. \citealt{Osaki74}). If the rotation is rapid enough, the growth time becomes much longer than the period of oscillation,  
to which we refer as overstable convective (OsC) mode. As shown numerically by \citet{LeeSaio86} for a $10\,M_\odot$ main-sequence model, an OsC mode can resonantly couple with a g mode in the surrounding radiative layers; in other words an envelope g mode can be excited by the OsC mode. The property of such resonant couplings are further discussed in \citet{LeeSaio87,LeeSaio89}.

The resonant excitation of g modes can occur in all fast-rotating early-type main-sequence stars, if the convective core rotates slightly faster than the radiative envelope. 
Those g modes should be observed with oscillation frequencies close to $|m|$ times the rotation frequency of the convective core, where $m$ is the azimuthal order of the oscillation.  This property may offer an explanation for the rotational modulations recently detected in many early type stars 
(e.g., \citealt{Balona19,Sikoraetal2019,Bowmanetal2018}) from $Kepler$ (e.g., \citealt{Boruckietal10}) and TESS (e.g., \citealt{Rickeretal15})  observations.

Rotational light modulation of a star appears in the Fourier spectrum as an isolated peak often accompanied by one or more harmonics. The cause of the modulation is usually attributed to the presence of a global magnetic field, while similar modulations are possible by chemically anomalous spots on the surface, ellipsoidal variations in close binaries, or eclipsing binaries (the latter two phenomena can be distinguished using other information).    
MOBSTER (Magnetic OB[A] stars with TESS) survey to find magnetic OB stars use the rotational light modulations to pre-select candidates for high-resolution spectro-polarimetric observations to detect magnetic fields (e.g., \citealt{David-Urazetal19}; \citealt{Sikoraetal2019}). 

\citet{Balona13,Balona17,Balona19} found about 20 to 40 percent of A and B stars to show rotational  light modulations. Attributing them to the presence of magnetic fields, \citet{Balona19} argued the need of gmajor revision of stellar physicsh because the dynamo to generate global magnetic fields should not work in the early type stars having no deep convection zones. However, this claim is at odds with the results of the surveys for magnetic OB stars that only 10 percent (or less) of them have magnetic fields, which are consistent with fossil fields (e.g., \citet{Wadeetal2014}; \citealt{Scholleretal2017}; \citealt{David-Urazetal20}).  In this paper we will show that rotational light modulations can be produced by g modes resonantly excited with over-stable convective modes in rotating convective cores  commonly present in early-type main-sequence stars.

Motivated with the importance of g modes resonantly excited by OsC modes in early type stars, we revisit the resonant excitation with improving the method of calculations.
Here, we make two improvements over the previous calculations.
One improvement concerns the number of expansion terms.
To represent the oscillation modes of rotating stars, \citet{LeeSaio86} employed series expansions of the perturbations in terms of spherical harmonic functions $Y_l^m(\theta,\phi)$,
but the number of $Y_l^m$s used for the expansions was very limited.  
We increase the number of expansion terms from 2 (\citealt{LeeSaio86}) to $\ge10$ so that
numerical results are insensitive to the number of expansion terms.
Second improvement concerns the superadiabatic temperature gradient in the convective core
\be
\epsilon\equiv \nabla-\nabla_{ad},
\ee
where $\nabla=d\ln T/d\ln p$ and 
$\nabla_{ad}=(\partial\ln T/\partial\ln p)_{ad}$ with $T$ and $p$ being the temperature and the pressure.
\citet{LeeSaio86} 
assumed $\epsilon=10^{-3}$
considering that a large value of $\epsilon$ would be required to keep efficient convective fluid motions in the rotating core. 
For a non-rotating convective core
mixing length theory of convection predicts $\epsilon$ to be as small as $10^{-7}$ to $10^{-6}$.
For rotating stars, however, 
we do not have any good knowledge of the magnitudes of $\epsilon$, except that
rotation would enhance $\epsilon$ by a large factor compared to that in non-rotating stars (e.g., \citealt{Stevenson79}).
In this paper, we employ $\epsilon=10^{-5}$ in most cases unless stated otherwise.

We carry out non-adiabatic calculations of 
unstable convective modes in the core of rotating $2M_\odot$ main sequence stars. 
We assume a weak differential rotation 
such that the convective core rotates slightly faster that the envelope \citep[e.g.][]{Lee88}.
The method of calculations of non-adiabatic oscillations of differentially rotating stars is the same as that used by 
\citet{LeeSaio93}. 
The numerical results for $2M_\odot$ main sequence models are given in \S 2. We discuss and conclude in \S 3 and \S 4.

\section{Numerical Analysis}

We express a pulsation mode in a rotating star by a sum of terms proportional to
spherical harmonic functions $Y_l^m(\theta,\phi)$
with different $l$s for a given $m$.
For example, the displacement vector $\pmb{\xi}$ is given as
\be
\xi_r(r,\theta,\phi,t)=r\sum_{j=1}^{j_{\rm max}}S_{l_j}(r)Y_{l_j}^m(\theta,\phi)e^{\rmi\sigma t},
\label{eq:xirexpand}
\ee
\be
\xi_\theta(r,\theta,\phi,t)=r\sum_{j=1}^{j_{\rm max}}\left[H_{l_j}(r){\partial\over\partial\theta}Y_{l_j}^m(\theta,\phi)
+T_{l'_j}{1\over\sin\theta}{\partial\over\partial\phi}Y_{l'_j}^m(\theta,\phi)\right]e^{\rmi\sigma t},
\ee
\be
\xi_\phi(r,\theta,\phi,t)=r\sum_{j=1}^{j_{\rm max}}\left[H_{l_j}(r){1\over\sin\theta}{\partial\over\partial\phi}Y_{l_j}^m(\theta,\phi)-T_{l'_j}{\partial\over\partial\theta}Y_{l'_j}^m(\theta,\phi)\right]e^{\rmi\sigma t},
\label{eq:xiphiexpand}
\ee
where $l_j=|m|+2(j-1)$ and $l'_j=l_j+1$ for even modes and $l_j=|m|+2j-1$ and $l'_j=l_j-1$ for odd modes
with $j=1,~2,~\cdots,~j_{\rm max}$\footnote{The terminology of even modes and odd modes used here
is the same as that in Lee \& Saio (1986).}.
The parameter $j_{\rm max}$ gives the length of expansions.
Substituting these expansions into linearized basic equations, we obtain a set of linear ordinary differential equations
for the expansion coefficients such as $S_l(r)$, $H_l(r)$, and so on \citep[e.g.,][]{LeeSaio86}.
The set of differential equations for non-adiabatic oscillation modes of differentially rotating stars
are given in \citet{LeeSaio93}.
We employ the Cowling approximation, neglecting the Euler perturbation of the gravitational potential. 
We also ignore the terms associated with centrifugal force, which is justified because most of the kinetic energy
of the convective and $g$-modes is confined into deep interior.
For expansions, we use $j_{\rm max}=10$ to 15, with which the frequencies and eigenfunctions become insensitive to $j_{\rm max}$.

\begin{figure}
\resizebox{0.4\columnwidth}{!}{
\includegraphics{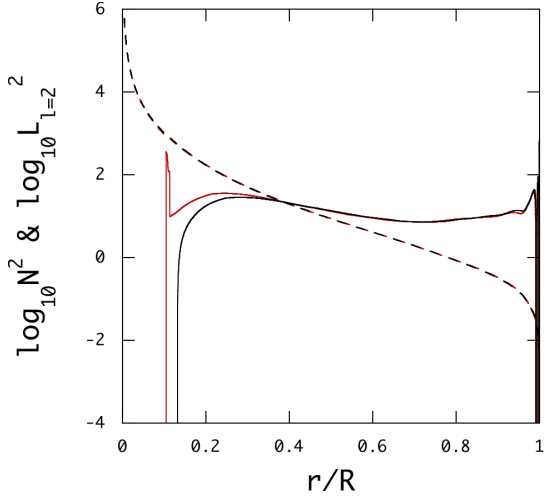}}
\caption{Propagation diagram for $2M_\odot$ main sequence models with $X_c=0.7$ (black lines) and $X_c=0.5$ (red lines) for the initial abundance $X=0.7$ and $Z=0.02$, 
where $X_c$ denotes the hydrogen abundance at the centre. Here, the solid lines are for the Brunt-V\"ais\"al\"a
frequency $N$ and the dashed lines for the Lamb frequency $L_l$ for $l=2$, respectively. These frequencies are normalized 
by $\sigma_0=\sqrt{GM/R^3}$.
}
\label{fig:propd}
\end{figure}

We use $2M_\odot$ main sequence star models with $X_c=0.7$ and $X_c=0.5$ as background models 
for modal analyses where $X_c$ is the hydrogen mass fraction at the stellar centre.
The models are computed by using a standard stellar evolution code with the OPAL opacity \citep{IglesiasRogers96} 
starting with the initial chemical composition $(X,Z)=(0.7, 0.02)$.
The stellar evolution code we use was originally written by \citet{Paczynski1970}, in which
the core evolution is followed from a chemically homogeneous model by a Henyey method with the outer boundary conditions provided by envelope models,
which are obtained simply by integrating the basic equations describing the static structure of stars
for given equations of state and opacity tables.
We applied the Schwarzschild criterion for convection and 
considered no mixing in the envelope.
Fig. \ref{fig:propd} shows the propagation diagrams of these models in which 
the Lamb frequency $L_l=\sqrt{l(l+1)}c/r$ and 
the Brunt-V\"ais\"al\"a frequency $N=\sqrt{-gA}$ are plotted versus the fractional radius $r/R$, 
where $c=\sqrt{\Gamma_1p/\rho}$ with $\Gamma_1=(\partial\ln p/\partial\ln\rho)_{ad}$ is the adiabatic sound velocity, and
$g=GM_r/r^2$ with $M_r=\int_0^r4\pi r^2\rho dr$ and $G$ the gravitational constant and 
$A=d\ln\rho/dr-\Gamma_1^{-1}d\ln p/dr$ is the Schwarzschild discriminant.
The frequencies $L_l$ and $N$ in the figure are normalized by $\sigma_0=\sqrt{GM/R^3}$ where $M$ and $R$ are the mass and radius of the stars.
Note that $N^2<0$ in the convective layers and $N^2>0$ in the radiative layers.
Each model has a convective core and subsurface thin convective layers in the
envelope.
In the slightly evolved model with $X_c=0.5$, the convective core has shrunk from 
the ZAMS model and is surrounded by a
$\mu$ gradient zone, which disturbs frequency spectra of $g$-modes propagating in the radiative envelope.
In the following, angular frequencies such as $\omega$ and $\Omega_s$ normalized by $\sigma_0$ are written as $\overline\omega$ and
$\overline\Omega_s$.

To compute unstable convective modes of rotating main sequence stars,
it is essential to set a non-zero super-adiabatic temperature in the convective core.
As mentioned in the previous section in most cases
we employ $\epsilon= 10^{-5}$, 
considering that $\epsilon$ 
should be enhanced by a large factor from that of non-rotating stars. 

As we discuss below, the oscillations in the envelope are affected by the presence of a differential rotation
between the convective core and radiative envelope.
We assume a differential rotation in some models using the formula given by
\be
\Omega(r)=\Omega_s\left[1+{b-1\over 1+e^{a(x-x_c)}}\right],
\label{eq:difrot}
\ee
where $x=r/R$ and $x_c$ corresponds to the core-envelope interface, $\Omega_s$ is the rotation speed at the stellar surface,
and $a$ and $b$ are parameters. 
Uniform rotation corresponds to $b=1$.
If $b>1$, 
the core rotates faster than the envelope.
In this paper we use $a=100$, for which $\Omega(r)$ stays $\approx b\Omega_s$ for $x<x_c$
but decreases steeply to $\Omega_s$ around $x_c$ (see an example in \citet{Lee88}).

For oscillation modes in differentially rotating stars, we use the symbol $\sigma$ for the angular frequency
(eigenfrequency) in the inertial frame.
Although the inertial frame frequency $\sigma$ does not depend on $r$, the frequency $\omega=\sigma+m\Omega(r)$ in
a local co-rotating frame depends on $r$ (but ${\rm Im}(\omega)={\rm Im}(\sigma)$).
If we let $\omega_c$ denote an oscillation frequency  
in the co-rotating frame of the core, 
the frequency $\omega_s$ in the co-rotating frame of the envelope is given by
\be
\omega_s=\omega_c-m(\Omega_c-\Omega_s)\approx \omega_c-m\Omega_s(b-1),
\label{eq:omega_env}
\ee
where $\Omega_c=\Omega(0)$.
If a prograde convective mode has a frequency $\omega_c>0$ for $m<0$,
the frequency $\omega_c$ should be shifted to  
$\omega_s$ in the envelope.
Then the $g$-mode in resonance with $\omega_s$ in the envelope should have a
radial order 
much lower than that of a $g$-mode having the frequency $\omega_c$ in the envelope.

As a tool to classify oscillation modes, we compute the fraction $f_j=E_j/E_K$ of energies where
$E_K$ and $E_j$ are the total and partial kinetic energies of an oscillation mode defined by
\be
E_K={1\over 2}\int_0^R r^4\omega_R^2\sum_{j=1}^{j_{\rm max}}
\left[|S_{l_j}|^2+l_j(l_j+1)|H_{l_j}|^2+l'_j(l'_j+1)|T_{l'_j}|^2\right]dr\equiv\sum_{j=1}^{j_{\rm max}}E_j,
\ee
and $\sum_jf_j=1$ where $\omega_{\rm R}$ is the real part of the frequency $\omega$.
In this paper, we pick up unstable convective modes that mainly consist of components 
with relatively slow latitudinal variations and satisfy 
\be
f_1+f_2+f_3\gtrsim 0.8.
\label{eq:f3cond}
\ee
The number 0.8 is somewhat arbitrary. 
If we use 0.9, instead of 0.8, the number of modes that satisfy the condition
will be smaller.

\subsection{Convective Modes in Uniformly Rotating Stars}

We start with the case of uniformly rotating stars for which $b=1$ and $\Omega=\Omega_c=\Omega_s$.
Adiabatic convective modes in the core of rotating stars appear as a pair of modes having frequencies 
$\omega_\pm$, 
and the relation $\omega_-=\omega_+^*$ holds.
Note that $\omega_\pm$ become pure imaginary when $\Omega=0$.
If we include non-adiabatic effects, core convective modes still appear as a pair of modes having $\hat\omega_\pm$,
but the relation $\hat\omega_-=\hat\omega_+^*$ holds only approximately even if the modes are well confined in the core
where non-adiabatic effects are insignificant.
In this paper, we discuss only the unstable modes for which imaginary part of frequency is negative.

\begin{figure}
\resizebox{0.33\columnwidth}{!}{
\includegraphics{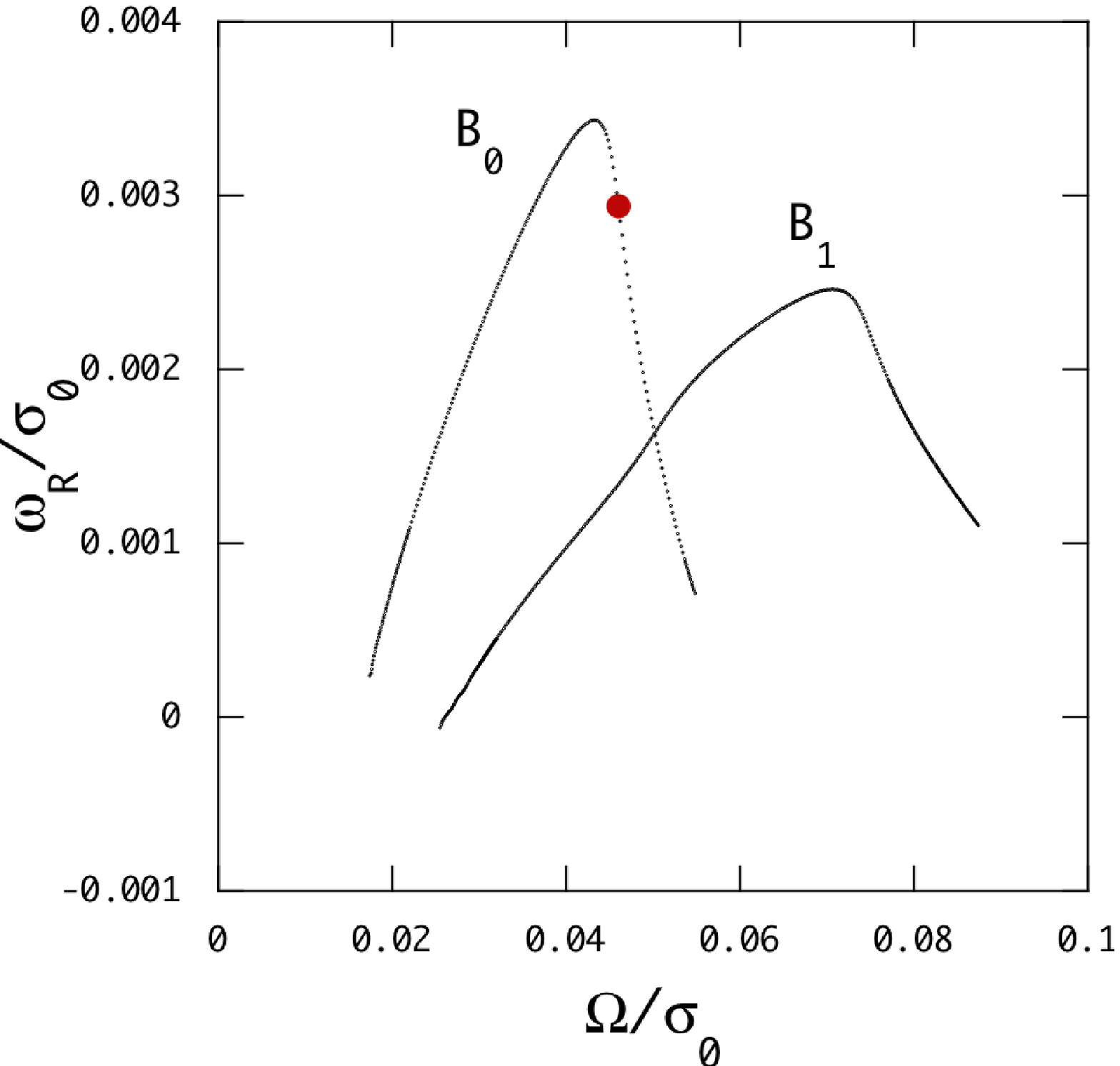}}
\resizebox{0.33\columnwidth}{!}{
\includegraphics{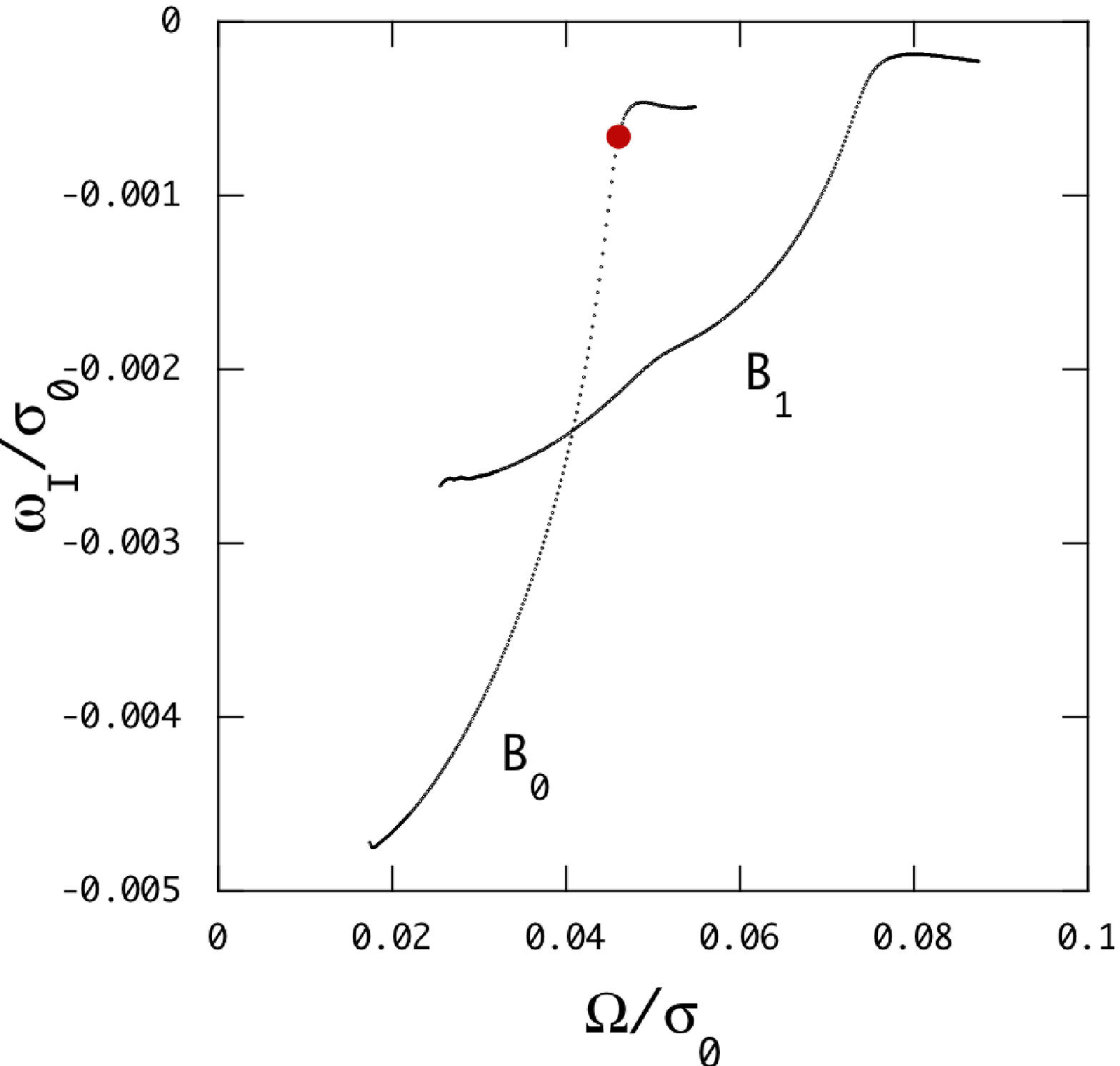}}
\caption{Complex eigenfrequency $\omega/\sigma_0$ of even-parity $m=-1$ unstable convective modes versus $\Omega/\sigma_0$ for the $2M_\odot$ ZAMS model for $b=1$ and $\epsilon=10^{-5}$ where $\omega\equiv\sigma+m\Omega$ is the oscillation frequency in the co-rotating frame, and $\omega_{\rm R}={\rm Re}(\omega)$ and $\omega_{\rm I}={\rm Im}(\omega)$.
Filled red circles indicate the eigenfrequency of the mode whose displacements in the interior are shown in Fig. \ref{fig:slhlm2md1b1}
below.
}
\label{fig:m2md1b1}
\end{figure}

The results of non-adiabatic calculations of even-parity $m=-1$ unstable prograde convective modes
for $\epsilon=10^{-5}$ are shown in Fig. \ref{fig:m2md1b1}, where normalized eigenfrequency in the co-rotating frame $\overline\omega=\overline\sigma+m\overline\Omega$ is plotted against
$\overline\Omega$. 
We have used $j_{\rm max}=10$ for the series expansion length.
Note that to obtain unstable convective modes for $\overline\Omega\not=0$ we do not calculate them 
from $\overline\Omega=0$ because there occurs frequent mode crossings as $\overline\Omega$ increases
and it is very difficult for us to compute convective modes from $\overline\Omega=0$ keeping their mode identity.
As shown in Fig.\ref{fig:m2md1b1}, the real part of the eigenfrequency, $\overline\omega_{\rm R}$, of a convective mode increases with $\overline\Omega$ up to a maximum  value and then decreases steeply. 
The absolute value of the imaginary part $|\overline\omega_{\rm I} |$ decreases steeply (i.e., getting stabilized) with increasing 
$\overline\Omega$, and stays almost constant for $\overline\Omega$ larger than the value at the peak of $\overline\omega_{\rm R}$.
We note that the rotation speed $\overline\Omega$ at the peak of $\overline\omega_{\rm R}$ for $B_0$ and $B_1$ modes is approximately proportional to $\sqrt{\epsilon}$. 
We stop calculating the modes at certain values of $\overline\Omega$ 
because the condition (\ref{eq:f3cond}) does not hold
any more as $\overline\Omega$ further increases.
Note that since $|\overline\omega_{\rm R}|\ll |m\overline\Omega|$ for convective modes, we have $\overline\sigma_{\rm R}\approx -m\overline\Omega$ in the inertial frame.

The unstable convective modes plotted in Fig. \ref{fig:m2md1b1} correspond to $B_n$ modes discussed by
\citet{Lee19} for rotating hot Jupiters.
\citet{Lee19} found two different kinds of convective modes, labeled $A_n$ and $B_n$ with $n$ being
the number of radial nodes of $S_{l_1}$ in the convective core.
Convective modes $A_n$ were first discussed by \citet{LeeSaio86} for a rotating $10M_\odot$ main sequence star,
for which $\epsilon=10^{-3}$ was assumed in the convective core.
The modal properties of unstable convective modes $A_n$ are reasonably well described by
using an asymptotic analysis based on the traditional approximation \citep[see, e.g.,][]{LeeSaio87,LeeSaio89,LeeSaio97}.
For rotating hot Jupiters, \citet{Lee19} found another kind of convective modes, labeled $B_n$, which need much larger
rotation speeds $\overline\Omega$ to be stabilized than $A_n$ convective modes.
For $\epsilon= 10^{-5}$, unstable convective modes $A_n$ 
are stabilized at much smaller values of $\overline\Omega$ and therefore only unstable convective modes
$B_n$ could survive for rapidly rotating stars.

\begin{figure}
\resizebox{0.33\columnwidth}{!}{
\includegraphics{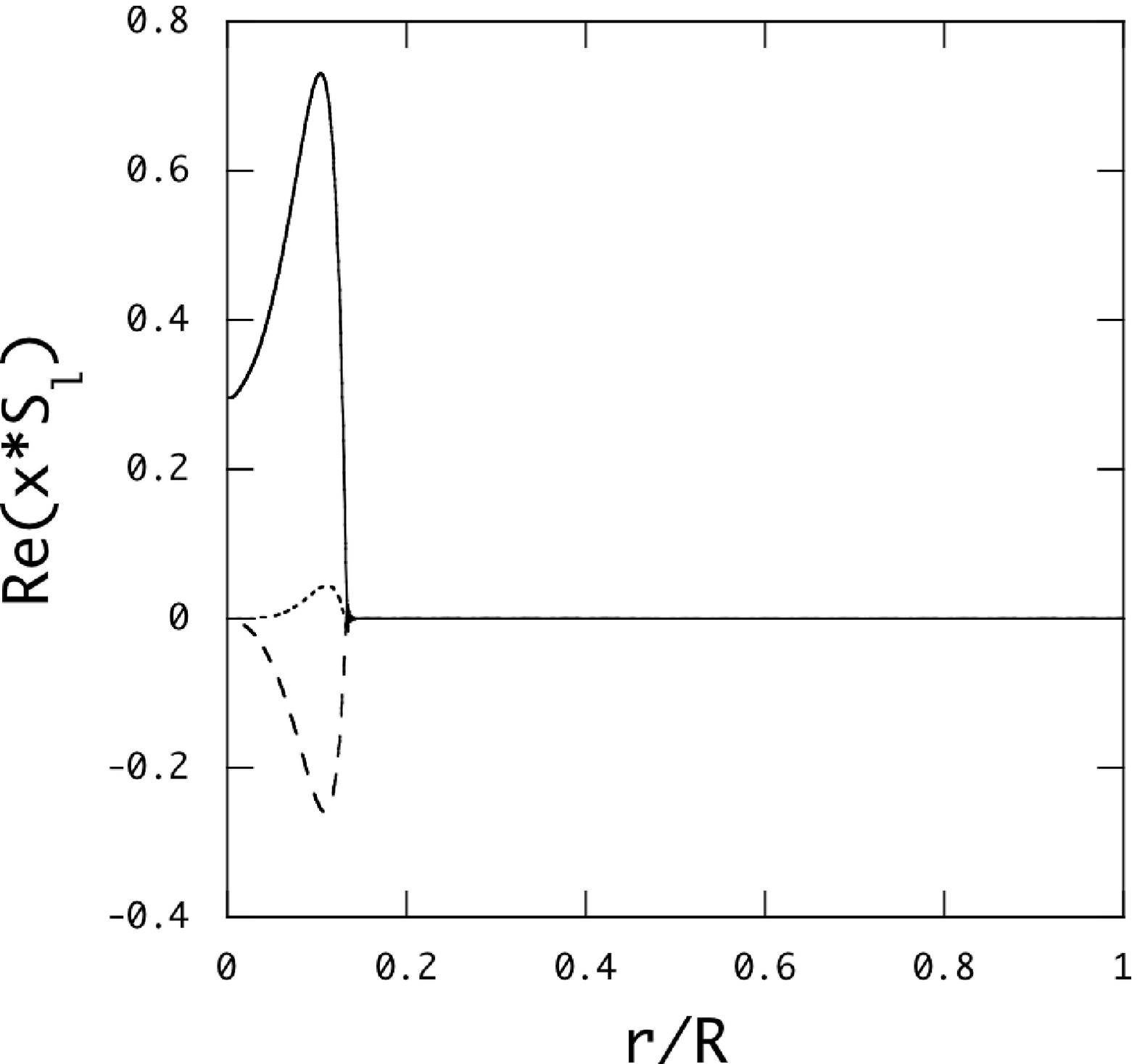}}
\resizebox{0.33\columnwidth}{!}{
\includegraphics{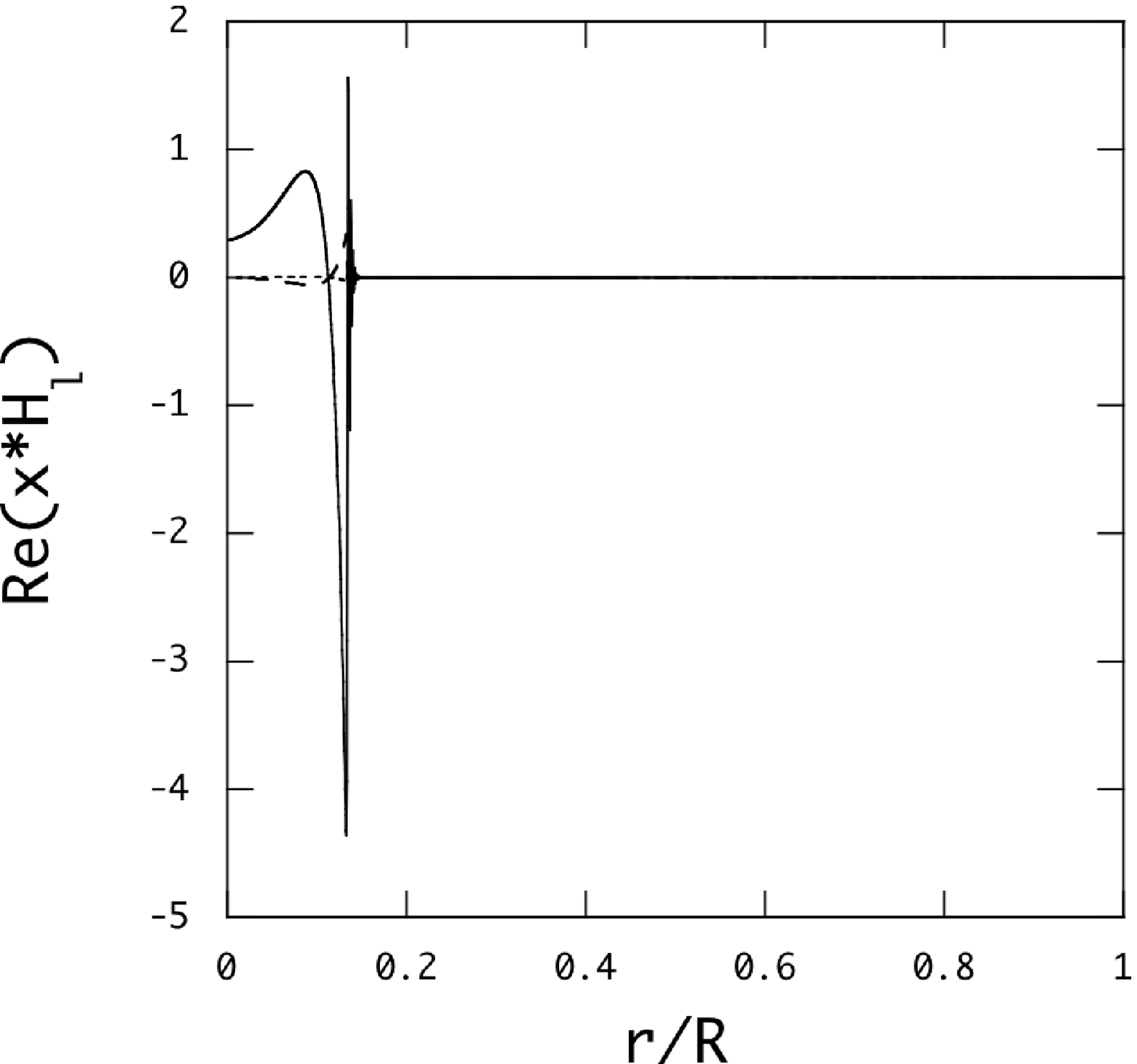}}
\caption{Real parts of the expansion coefficients for radial and horizontal displacements $xS_l$ and $xH_l$ as a function of $x=r/R$ for the $m=-1$ $B_0$ mode
with $\overline\omega=(2.94\times10^{-3}, -6.59\times10^{-4})$ at $\overline\Omega=0.046$ (indicated by filled red circles in Fig. \ref{fig:m2md1b1}). 
Here, the solid, dashed, and dotted lines represent the coefficients with $l=1$, 3, and 5, respectively.
The amplitudes are normalized by the maximum value of $|xS_l|$ in the interior.
}
\label{fig:slhlm2md1b1}
\end{figure}

Fig. \ref{fig:slhlm2md1b1} shows the first few expansion coefficients for radial and
horizontal displacements, $xS_l$ and $xH_l$, of 
the $m=-1$ prograde convective mode with $\overline\omega=(2.94\times10^{-3}, -6.59\times10^{-4})$ at $\overline\Omega=0.046$,
which is the $B_0$ mode indicated by filled red circles in Fig. \ref{fig:m2md1b1}.
Note that this mode has small $|\overline\omega_{\rm I}|$ expected for coupling with envelope g-modes
and still satisfies the condition (\ref{eq:f3cond}).
The amplitudes of $S_l$ and $H_l$ are completely confined in the convective core and no amplitude penetration
into the radiative envelope takes place.
We note that because of the small frequency $\overline\omega_{\rm R}\sim 3\times10^{-3}$, 
envelope $g$-modes corresponding to this frequency are extremely high radial order modes having radial nodes
more than 600.

We have carried out similar calculations for the slightly evolved model with $X_c=0.5$ and have confirmed that
no amplitude penetration of convective modes into the envelope takes place for uniform rotation.

\subsection{Cases with Weak Differential Rotation}

In the previous section we found that 
no amplitude penetration of convective modes into the envelope can be expected in uniformly rotating stars.
In this subsection we discuss convective modes in models with a weak differential rotation given by equation (\ref{eq:difrot})
with $b=1.2$.
As equation  (\ref{eq:omega_env}) indicates, if the convective core rotates slightly faster than the envelope,
a $g$-mode which couples with a convective mode should have a higher frequency and
hence a smaller number of nodes in the envelope.
We expect such a $g$-mode to get less dissipation and to have larger amplitude in the envelope.

\subsubsection{ZAMS model}

\begin{figure}
\resizebox{0.33\columnwidth}{!}{
\includegraphics{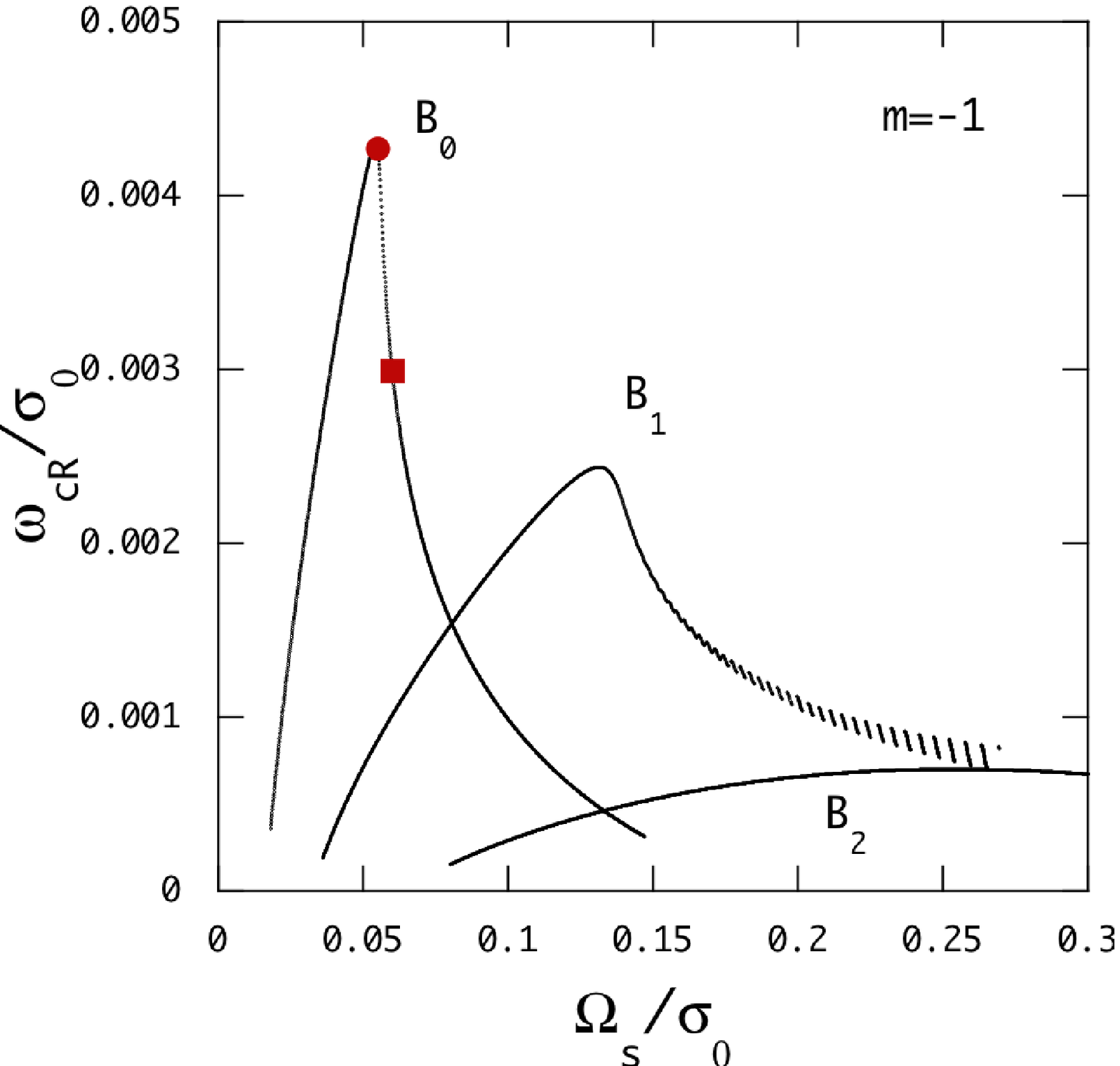}}
\resizebox{0.33\columnwidth}{!}{
\includegraphics{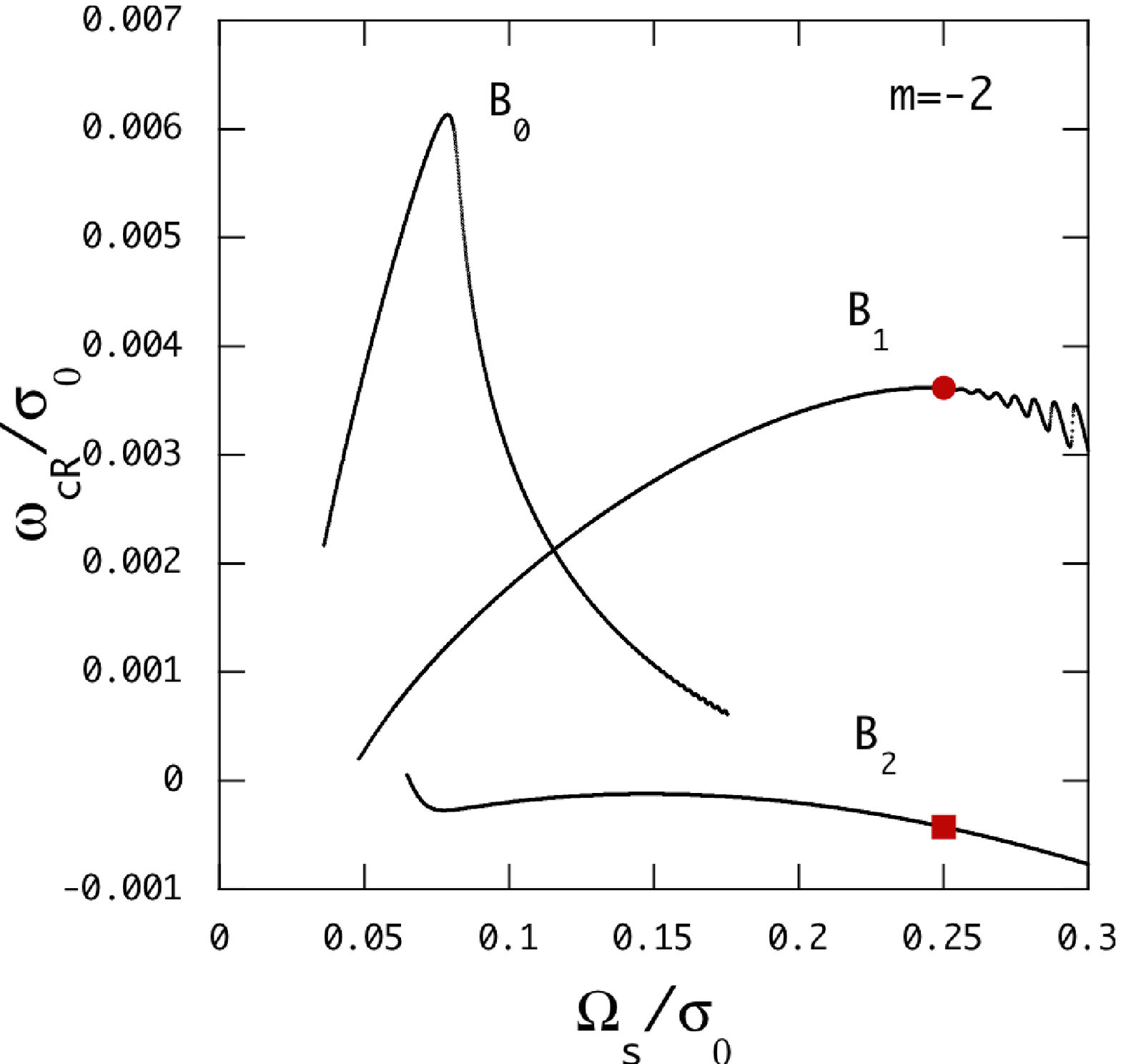}}
\resizebox{0.33\columnwidth}{!}{
\includegraphics{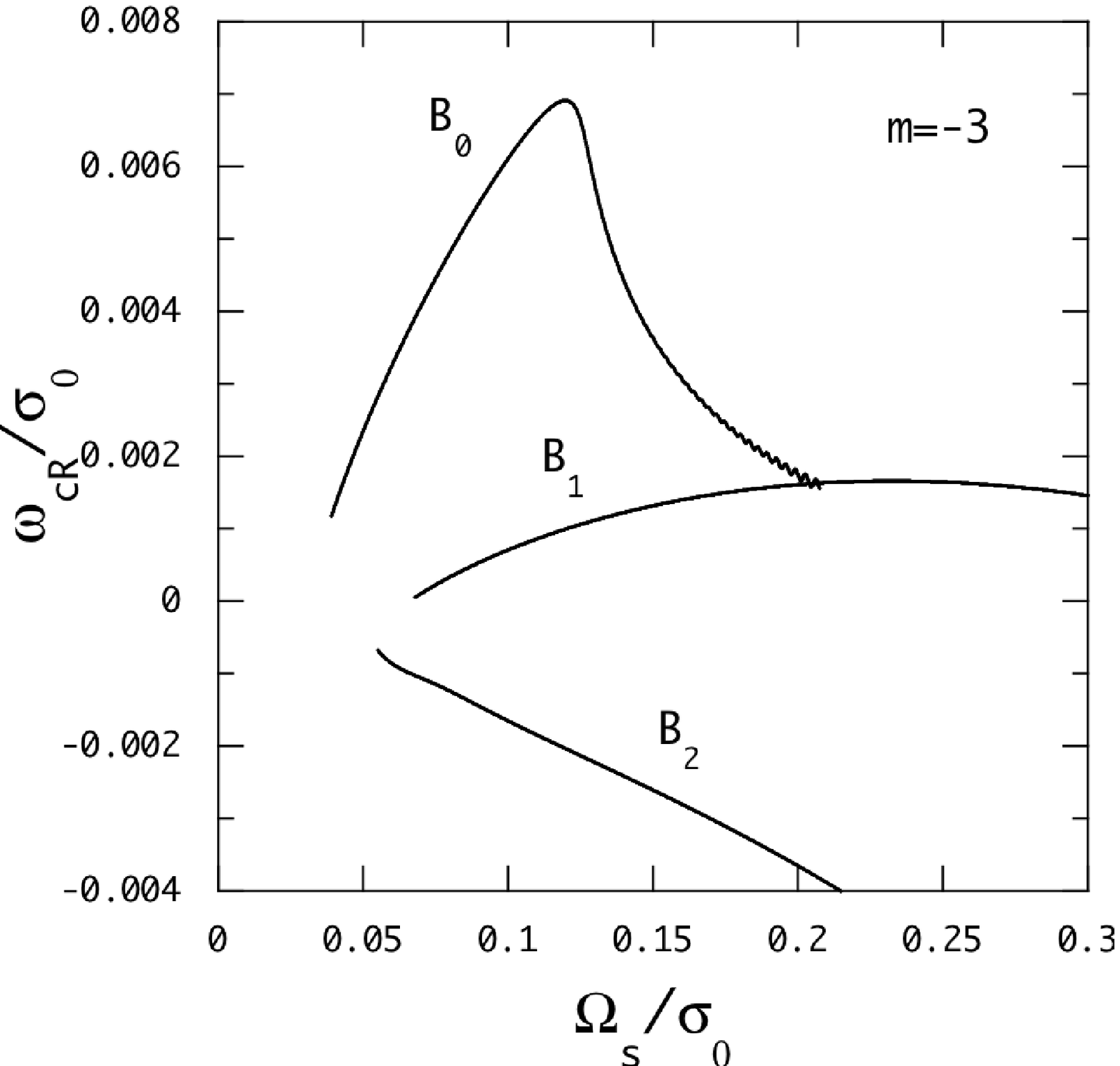}}
\resizebox{0.33\columnwidth}{!}{
\includegraphics{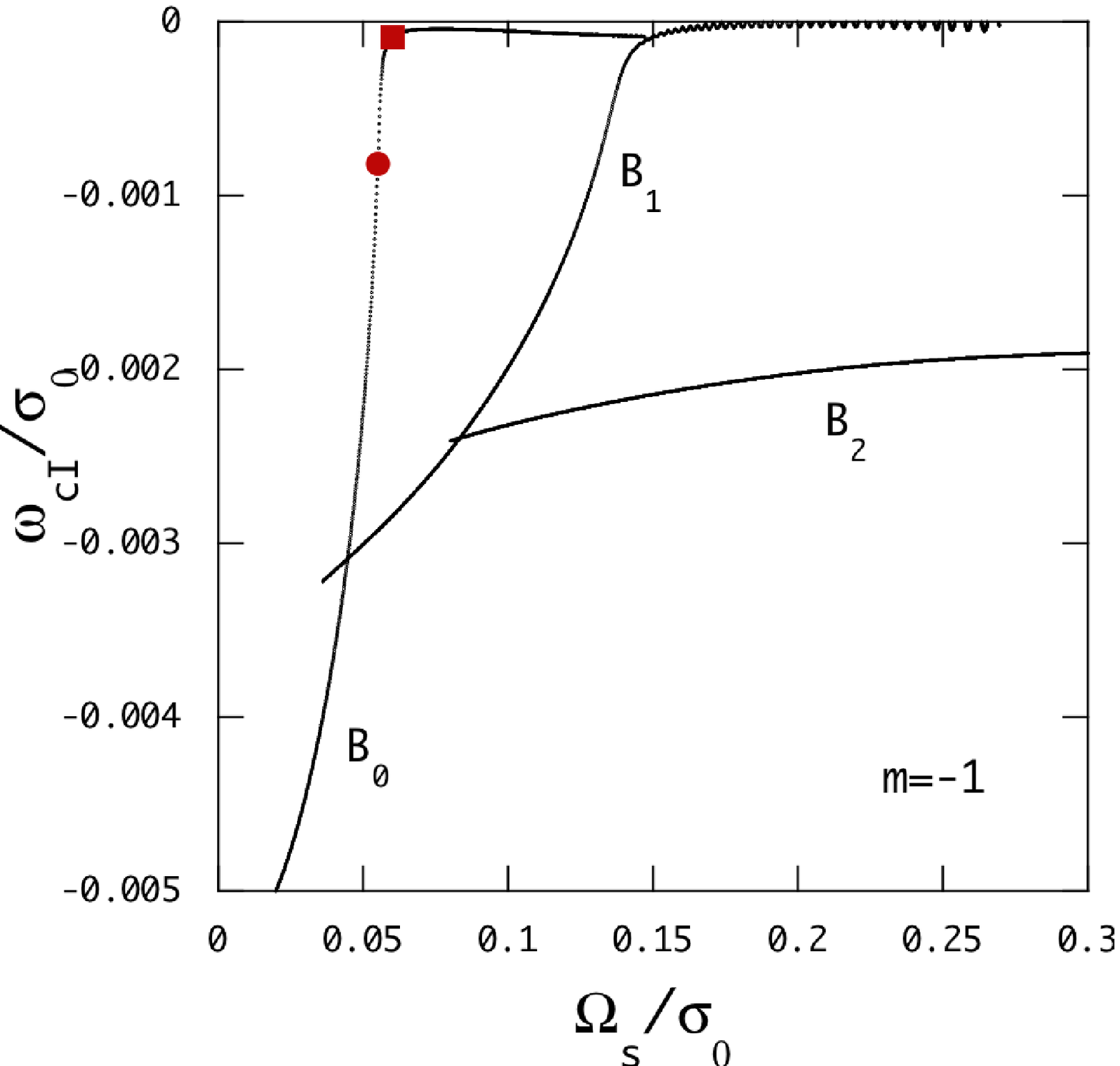}}
\resizebox{0.33\columnwidth}{!}{
\includegraphics{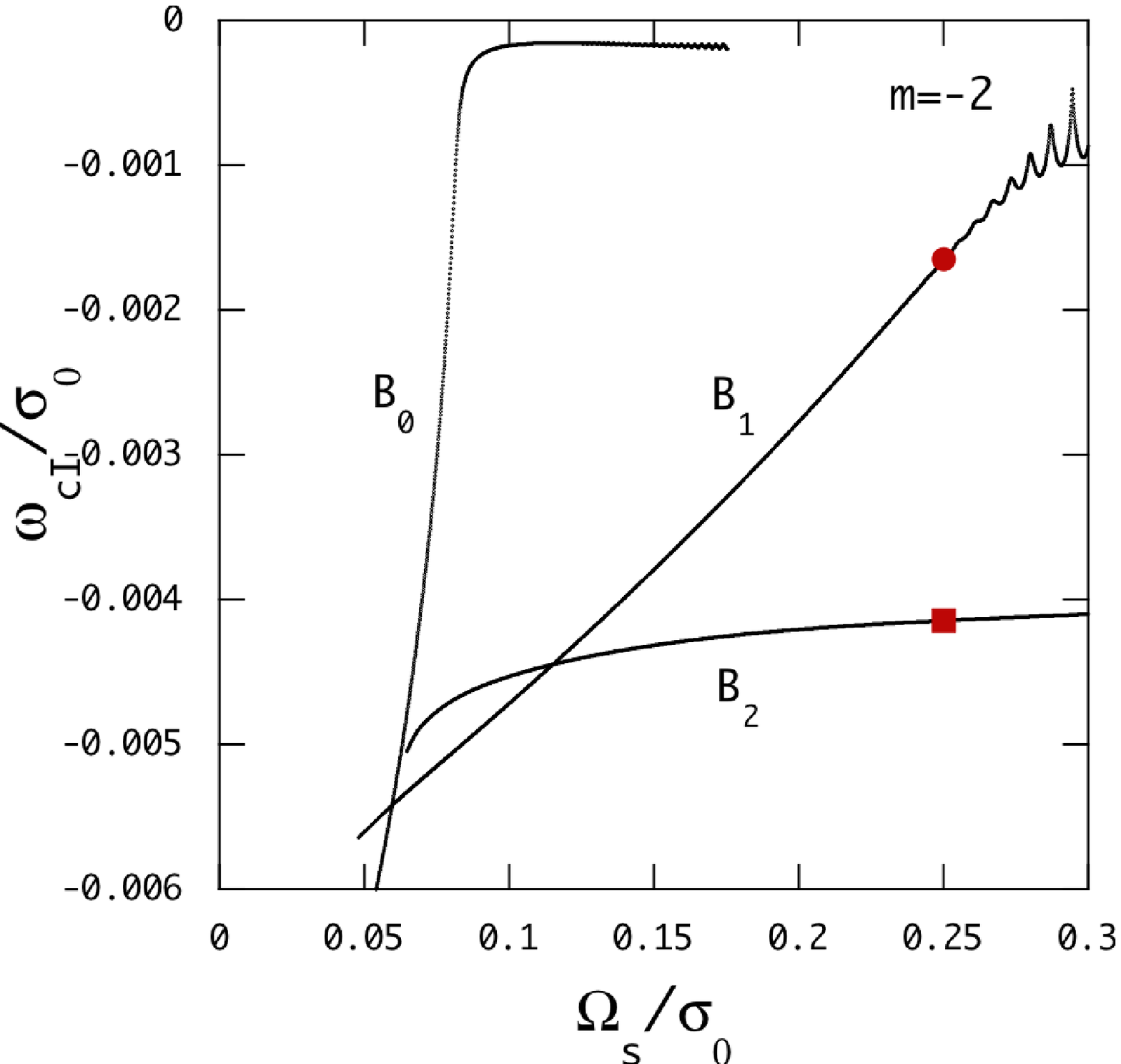}}
\resizebox{0.33\columnwidth}{!}{
\includegraphics{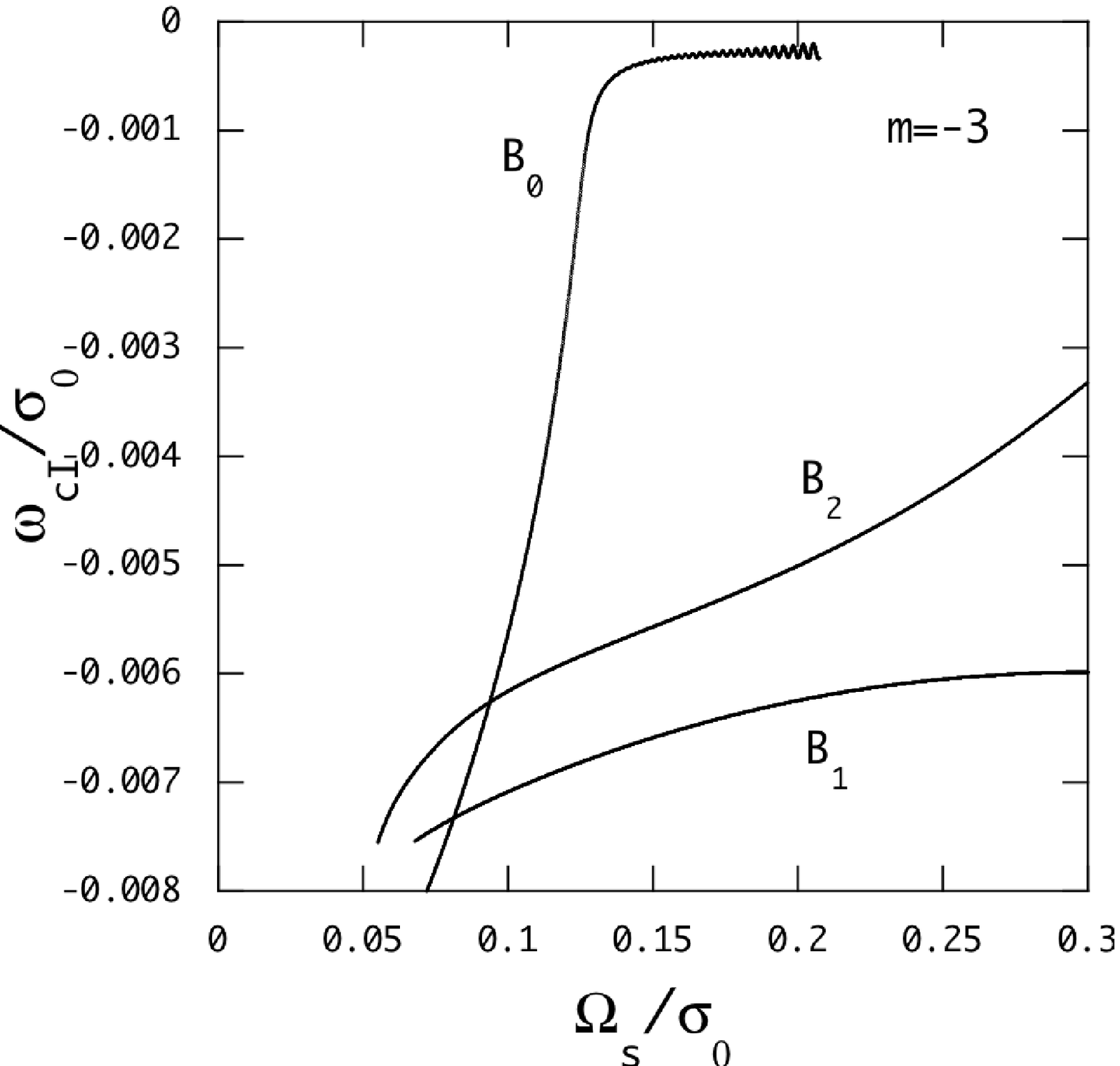}}
\caption{Complex eigenfrequency $\omega_c/\sigma_0$ of even-parity convective modes of $m=-1$, $-2$, and $-3$ 
versus $\Omega_s/\sigma_0$ for the differentially rotating $2M_\odot$ ZAMS model with $b=1.2$ and $\epsilon=10^{-5}$ where 
$\omega_c\equiv\sigma+m\Omega_c$ with
$\Omega_c$ being the rotation speed at the centre, and $\omega_{c{\rm R}}={\rm Re}(\omega_c)$ and $\omega_{c{\rm I}}={\rm Im}(\omega_c)$.
Filled red circles and squares indicate the eigenfrequencies of the modes whose displacements in the interior
are shown in Figs. 5 to 8. 
Note that only unstable modes having $\omega_{c{\rm I}}<0$ are plotted.
}
\label{fig:m2md1b1p2}
\end{figure}

For the $2M_\odot$ zero age main sequence (ZAMS) model with $\epsilon=10^{-5}$ and $b=1.2$, we have 
obtained unstable convective modes of even parity for $m=-1$, $-2$, and $-3$.
The results 
are shown in Fig. \ref{fig:m2md1b1p2}, in which the complex frequency
$\overline\omega_c$ is plotted versus $\overline\Omega_s$ for $B_0$, $B_1$, and $B_2$ modes,  where 
$\omega_c$ is the frequency in the frame co-rotating at the centre given as
$\omega_c=\sigma+m\Omega_c$ with $\Omega_c=\Omega(0)$, and $\omega_{c{\rm R}}={\rm Re}(\omega_c)$ and $\omega_{c{\rm I}}={\rm Im}(\omega_c)$.

As in the case of uniformly rotating stars, 
$\overline\omega_{c{\rm R}}$ of an unstable convective mode has a maximum and $|\overline\omega_{\rm I}|$ decreases with increasing $\overline\Omega_s$
and at the maximum of $\overline\omega_{\rm R}$ 
we find $|{\rm Im}(\overline\omega_c)|\ll{\rm Re}(\overline\omega_c)$, that is, 
the convective mode is stabilized to have a very small growth rate.
\footnote{For differentially rotating stars, however, there occur some exceptions to this rule, for example,
for the convective modes that have a co-rotation point at which $\omega_{\rm R}=0$ in the interior.
In addition, some convective modes which are stabilized only by rapid rotation have
no sharp maximum peak of $\overline\omega_{c{\rm R}}$. }
As $\overline\Omega_s$ further increases 
both $\overline\omega_{c{\rm R}}$ and $\overline\omega_{c{\rm I}}$ show rapid variations with small amplitudes
as a function of $\overline\Omega_s$, which are caused by  
resonances with envelope $g$-modes
having a dense frequency spectrum. 
The amplitudes and separation of this resonance feature, as clearly seen for the $B_1$ modes, increase 
as $\overline\Omega_s$ increases.
This corresponds to the fact that the frequency $\overline\omega_s\approx \overline\omega_c-m(\overline\Omega_c-\overline\Omega_s)$  in the envelope increases
and hence the frequency separation  
of $g$-modes increases with increasing $\overline\Omega_s$.
We note that $\overline\omega_{c{\rm I}}$ of convective modes 
remains negative with increasing $\overline\Omega_s$ beyond the value of $\overline\Omega_s$ at the $\overline\omega_{c{\rm R}}$ maximum, 
except for the $m=-1$ $B_1$ mode, whose $\overline\omega_{c{\rm I}}$ frequently changes its sign
as $\overline\Omega_s$ increases.
In general, we find it difficult to obtain accurate eigenfrequencies of convective modes when both 
$|\overline\omega_{c{\rm R}}|$ and $|\overline\omega_{c{\rm I}}|$ are very small, and hence we have to stop computing convective modes at certain $\overline\Omega_s$.

It is interesting to note that $\overline\omega_{c{\rm R}}$ of $B_2$ modes of $m=-2$ and $-3$ become negative
for $\overline\Omega_s\gtrsim 0.05$.
These modes are retrograde to the rotation in the convective core and have a co-rotation point
at which $\overline\omega_{\rm R}=0$ near the convective core boundary.

For a given $m$, 
the rotation rate $\overline\Omega_s$ at the peak of $\overline\omega_{c{\rm R}}$ for $B_1$ mode is larger than that 
for $B_0$ mode, and the height of the peak for $B_1$ mode is lower than that for $B_0$ mode.
If we compare the results for $m=-1$ and $m=-2$,
$\overline\Omega_s$ at the $\overline\omega_{c{\rm R}}$ peak for the $B_0$ mode is smaller for $m=-1$ than for $m=-2$, that is,
we need more rapid rotation to stabilize unstable convective modes of $m=-2$ than those of $m=-1$.

\begin{figure}
\resizebox{0.33\columnwidth}{!}{
\includegraphics{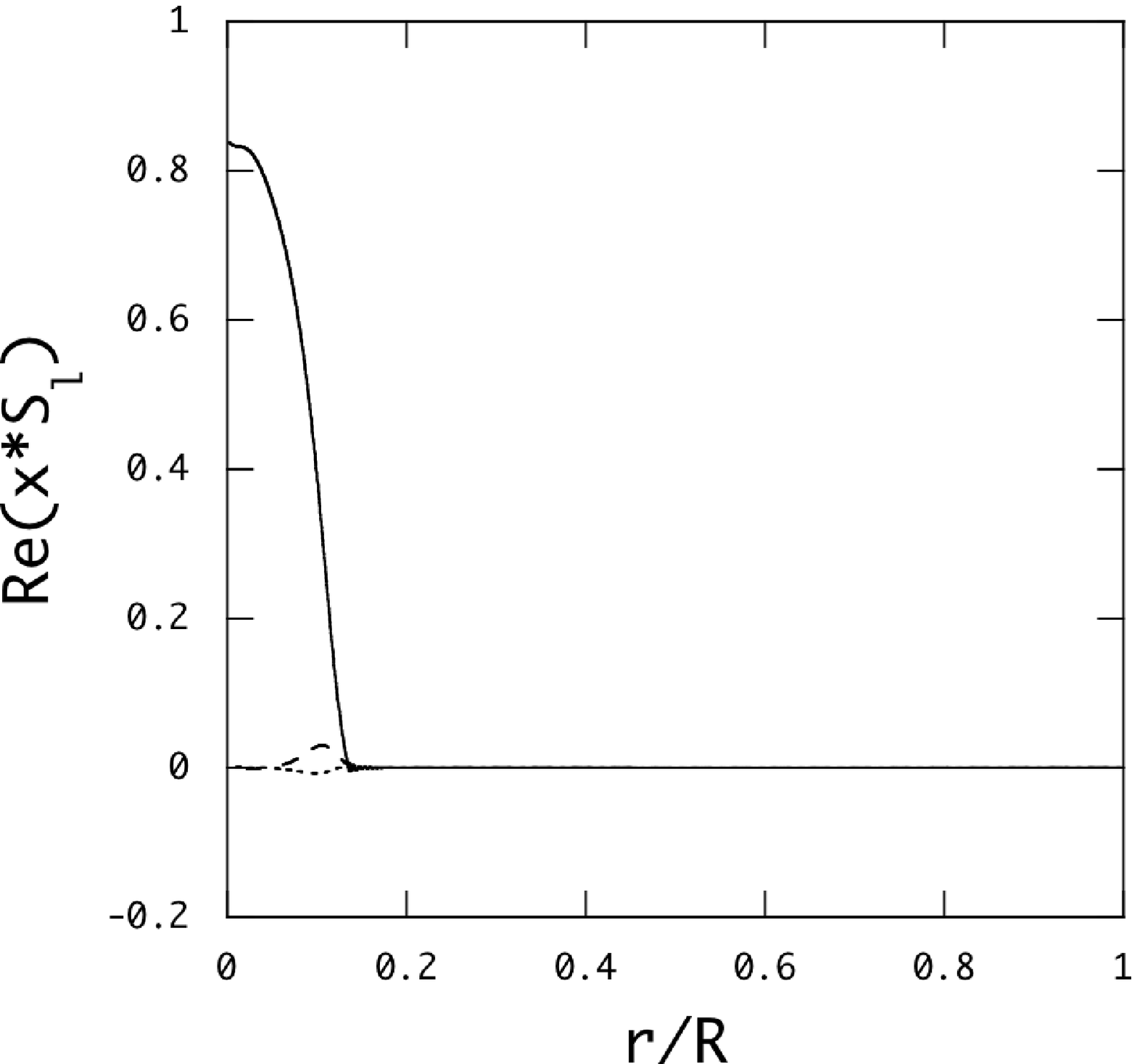}}
\resizebox{0.33\columnwidth}{!}{
\includegraphics{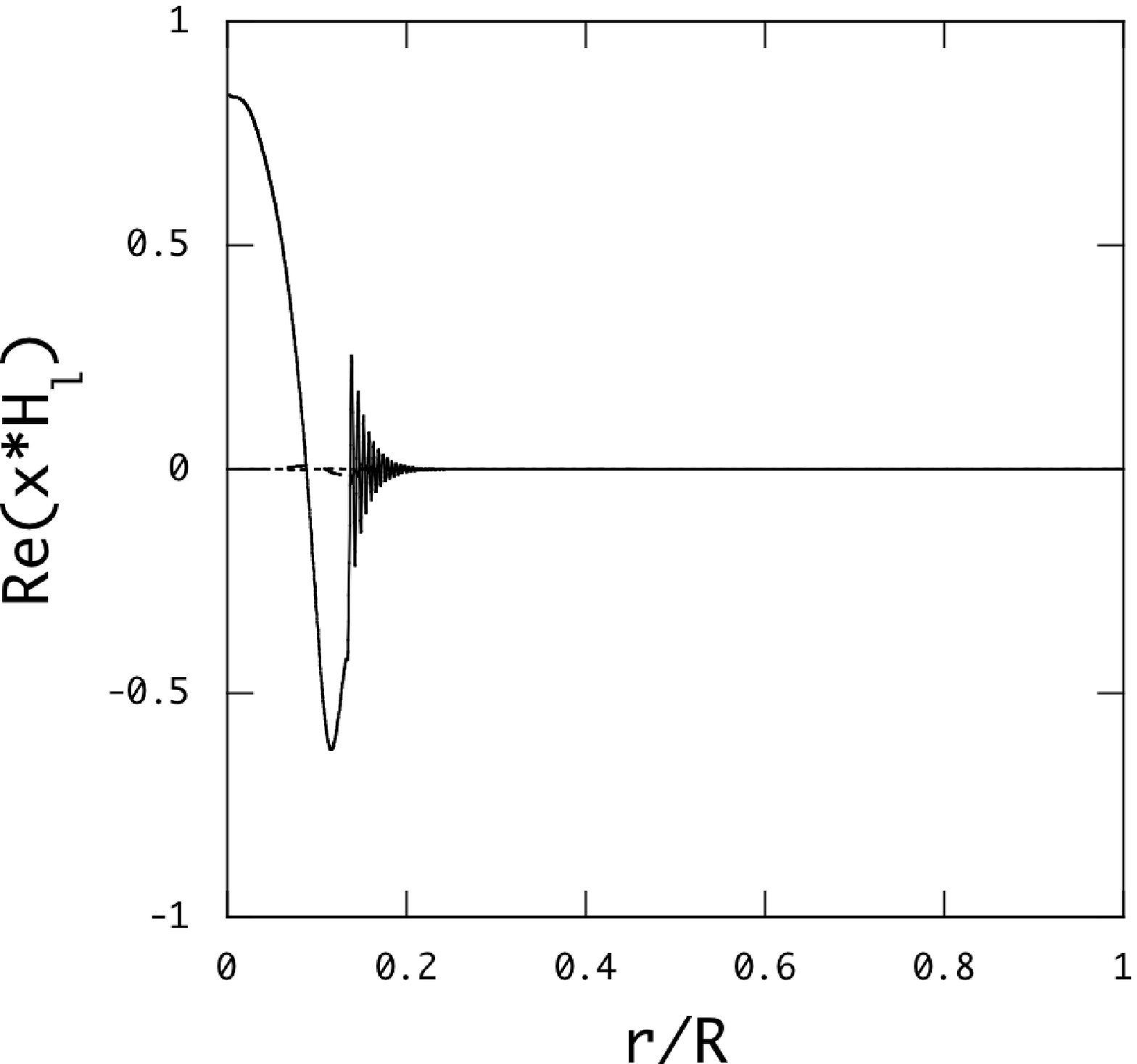}}
\caption{Real parts of the expansion coefficients for radial and horizontal displacements $xS_l$ and $xH_l$ as a function of $x=r/R$ for the $m=-1$ $B_0$ mode
with $\overline\omega_c=(4.27\times10^{-3}, -8.18\times10^{-4})$ at $\overline\Omega_s=0.055$ (indicated by filled red circles in the left panels of
Fig. \ref{fig:m2md1b1p2}). 
Here, 
the solid, dashed, and dotted lines represent the coefficients with $l=1$, 3, and 5, respectively. 
}
\label{fig:slhl_m2md1b1p2mm1_0055}
\end{figure}

\begin{figure}
\resizebox{0.33\columnwidth}{!}{
\includegraphics{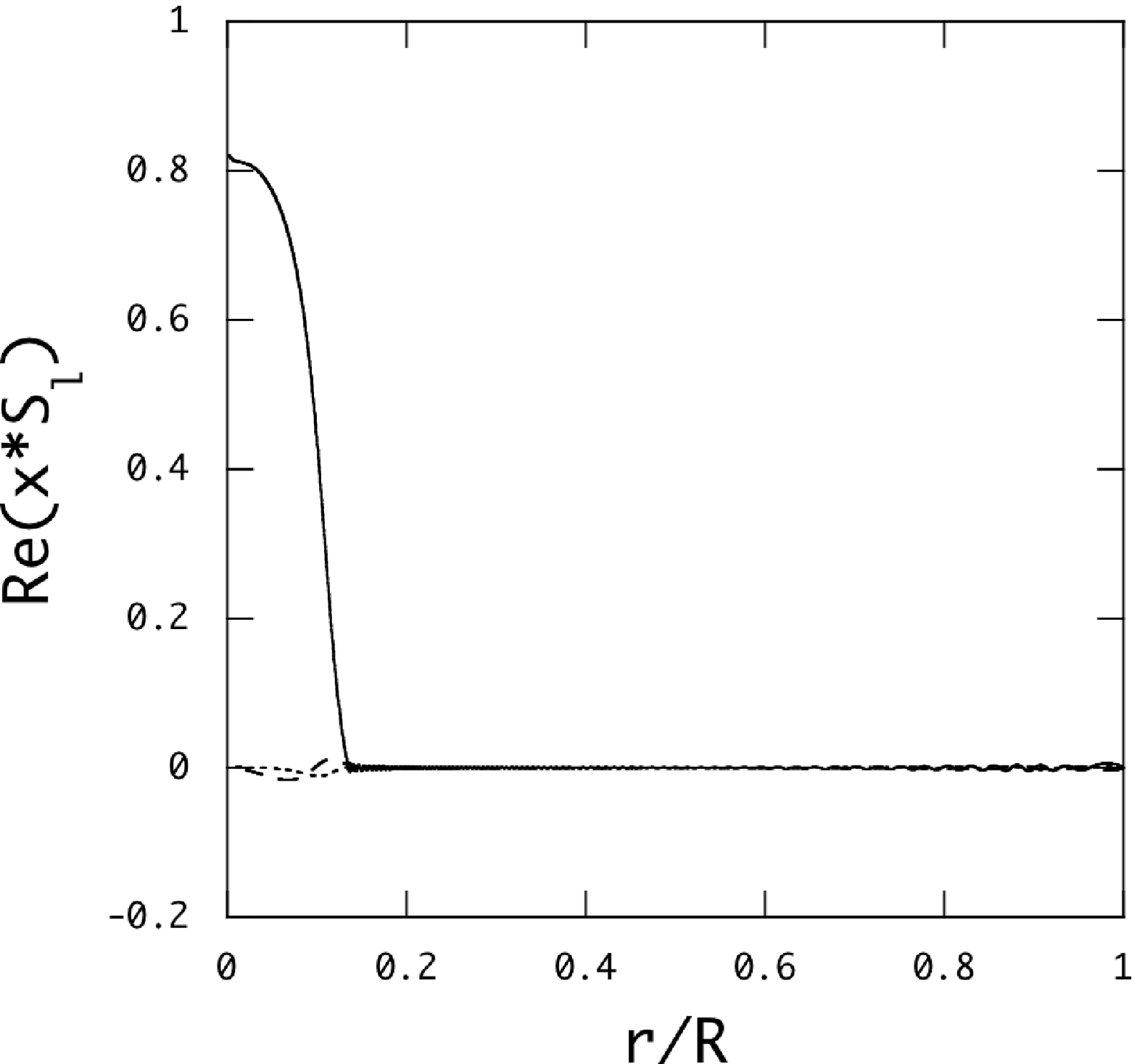}}
\resizebox{0.33\columnwidth}{!}{
\includegraphics{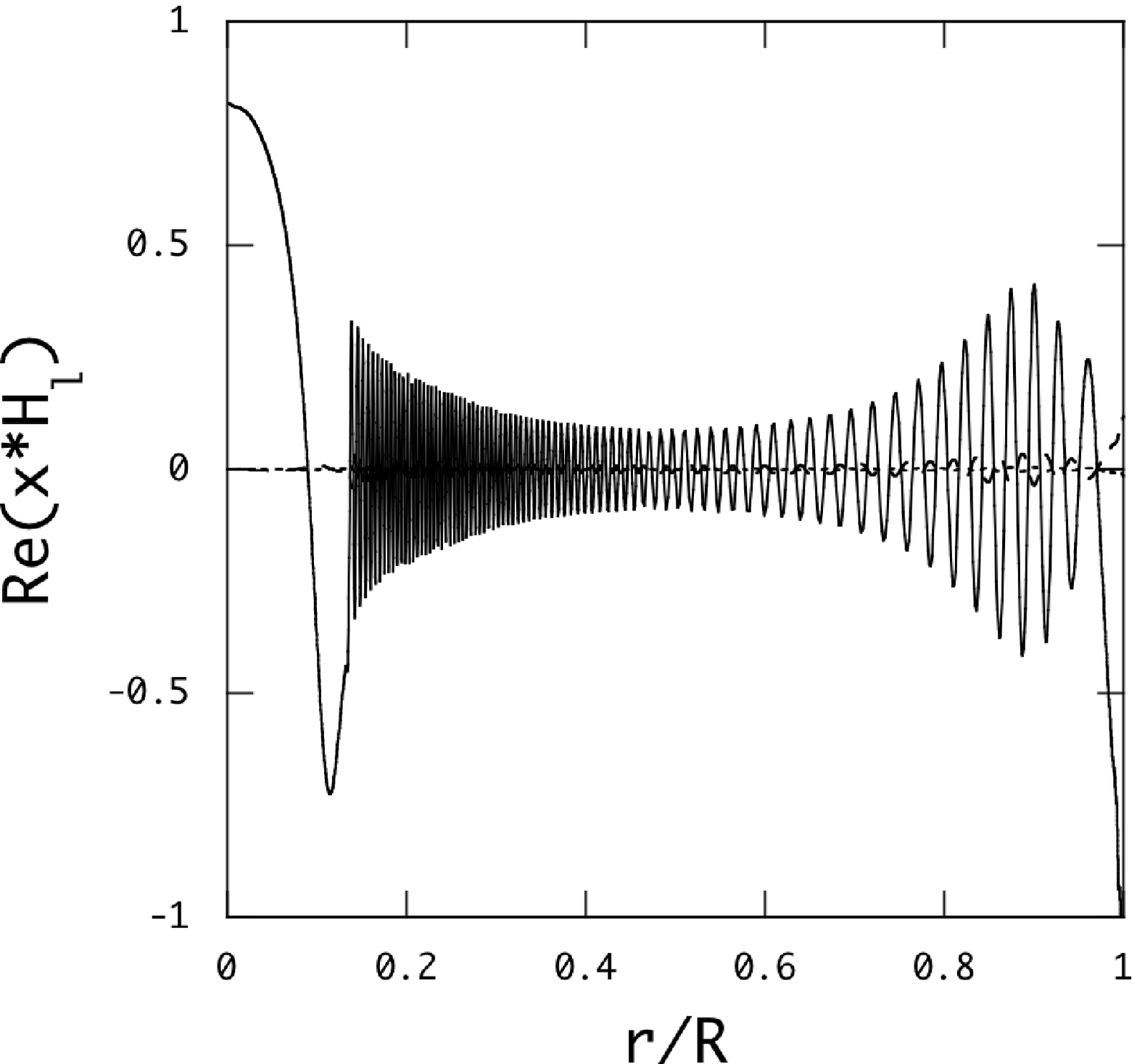}}
\caption{Same as Fig. \ref{fig:slhl_m2md1b1p2mm1_0055} but for $\overline\omega_c=(2.99\times10^{-3}, -8.84\times10^{-5})$
at $\overline\Omega_s=0.06$ (indicated by filled red squares in the left panels of Fig. \ref{fig:m2md1b1p2}).
}
\label{fig:slhl_m2md1b1p2mm1_006}
\end{figure}

We find that the amplitudes of unstable convective modes can 
penetrate deep into the envelope, even reach to the stellar surface, when the convective modes are stabilized to have
sufficiently small $|\overline\omega_{c{\rm I}}|$ (i.e., long growth time).
This amplitude penetration into the envelope takes place as a result of resonance
with an envelope $g$-mode. 
Examples of such amplitude penetration of unstable convective modes
are shown in Figs. \ref{fig:slhl_m2md1b1p2mm1_0055} and
\ref{fig:slhl_m2md1b1p2mm1_006}, where the real parts of the expansion coefficients for radial and horizontal displacements $xS_l$ and $xH_l$ are plotted as a function of $x=r/R$
for the $m=-1$ $B_0$ mode at two different rotation rates $\overline\Omega_s=0.055$ and 0.06.
As shown in these figures,
the first expansion coefficients $S_l$ and $H_l$ with $l=|m|$ are dominating 
over other expansion components both in the core and in the envelope, suggesting $f_{j>1}\ll f_{j=1}\cong 1$ 
(see equation (\ref{eq:f3cond})).
For the convective $B_0$ mode, 
the first component $S_{l=|m|}$ has no radial nodes in the core although it has numerous nodes in the envelope, that is,
the convective mode is coupled with a high radial order $g$-mode in the envelope.
Fig. \ref{fig:slhl_m2md1b1p2mm1_0055} for $\overline\Omega=0.055$ where $\overline\omega_{c{\rm R}}$
is close to the maximum shows only weak penetration into the envelope.
The amplitudes of $H_l$ rapidly decays within a thin layer exterior to 
the convective core.
On the other hand, Fig. \ref{fig:slhl_m2md1b1p2mm1_006} shows an efficient penetration of the amplitudes into
the envelope, reaching to the stellar surface, which is clearly seen particularly for the coefficients $H_l$, 
because horizontal displacements of high order $g$-modes are much larger than radial displacements by
$\sim 1/|\overline\omega_s|^2$.
We find that the number of radial nodes in the envelope amounts to $\sim 100$ for the mode shown in the figure.

\begin{figure}
\resizebox{0.33\columnwidth}{!}{
\includegraphics{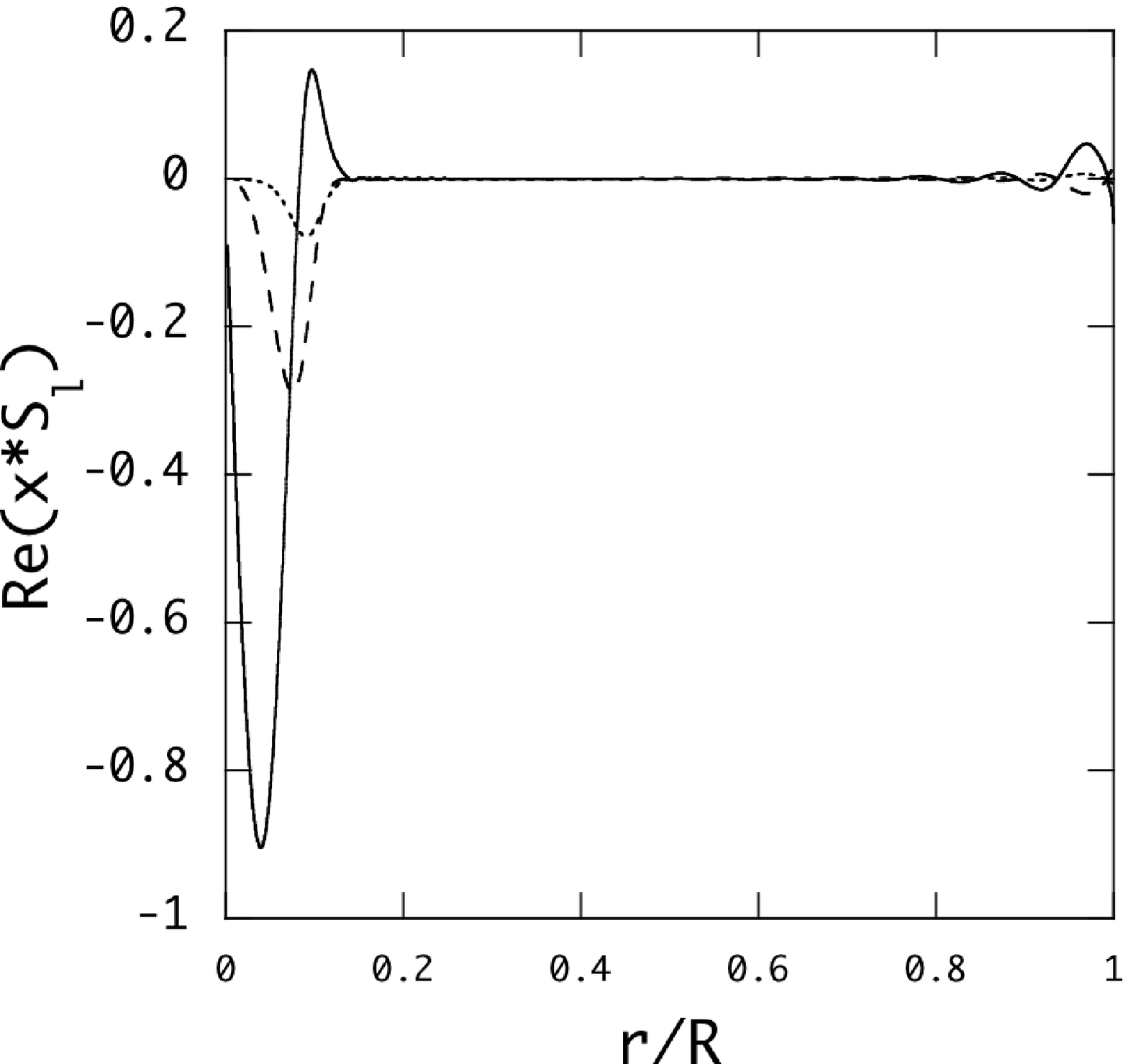}}
\resizebox{0.33\columnwidth}{!}{
\includegraphics{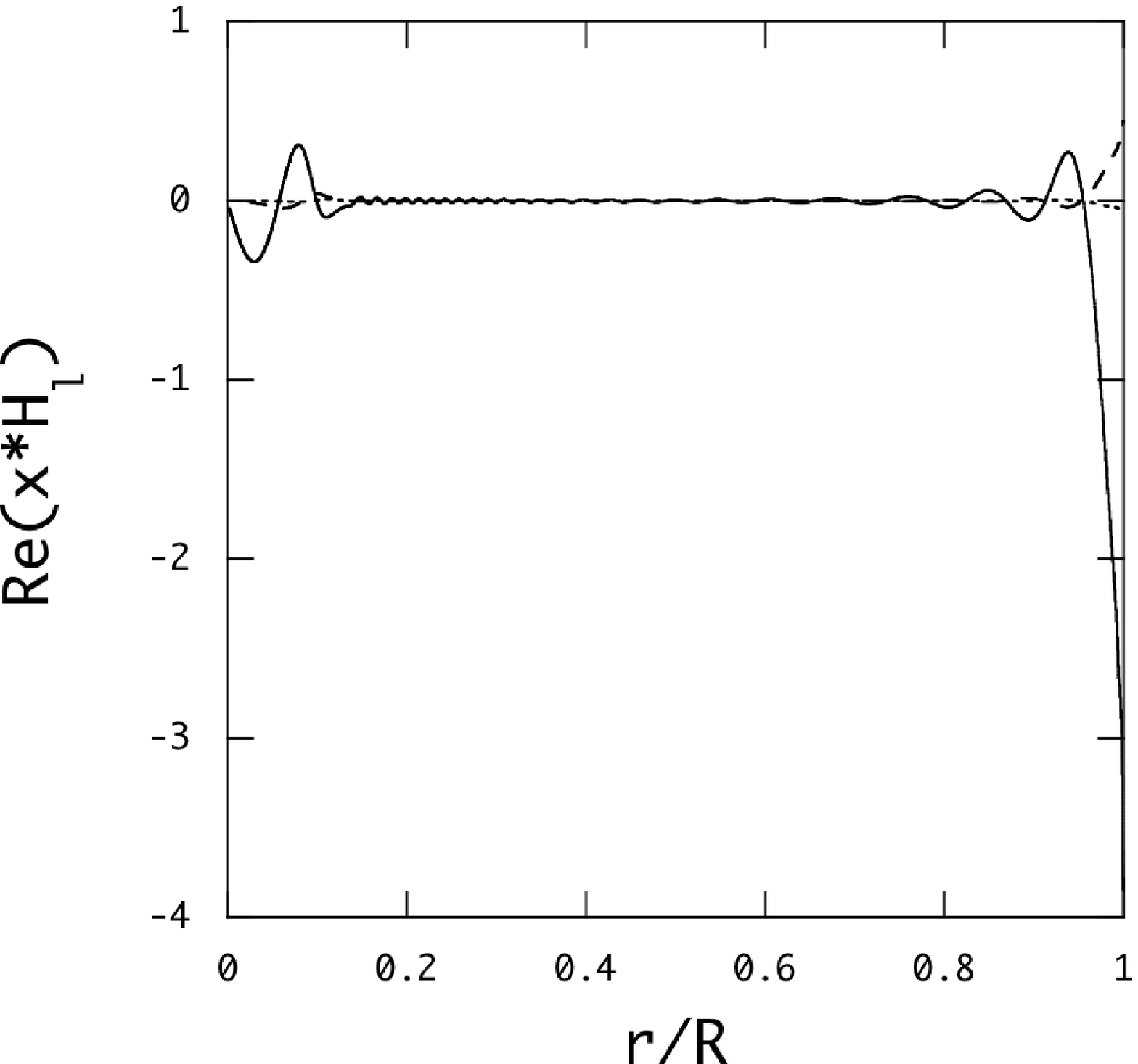}}
\caption{Real parts of the expansion coefficients for radial and horizontal displacements $xS_l$ and $xH_l$ as a function of $x=r/R$ 
for the $m=-2$ $B_1$ mode 
with $\overline\omega_c=(3.63\times10^{-3}, -1.49\times10^{-3})$ at $\overline\Omega_s=0.25$ (indicated by filled red circles in the middle panels of Fig. \ref{fig:m2md1b1p2}). 
the solid, dashed, and dotted lines represent the coefficients with $l=2$, 4, and 6, respectively. 
}
\label{fig:slhl_m2md1b1p2mm2_025}
\end{figure}

Fig. \ref{fig:slhl_m2md1b1p2mm2_025} is the same as Fig. \ref{fig:slhl_m2md1b1p2mm1_0055}
but for the $m=-2$ $B_1$ mode with 
$\overline\omega_c=(3.63\times10^{-3}, -1.49\times10^{-3})$ at $\overline\Omega_s=0.25$
(shown by filled red circles in the middle panels in Fig. \ref{fig:m2md1b1p2}).
As seen in this figure, $S_{l=|m|}$ has one node in the core, and the coefficients $S_{l=|m|+2}$ and $S_{l=|m|+4}$
have noticeable amplitudes there.
Fig. \ref{fig:slhl_m2md1b1p2mm2_025} shows that $B_1$ mode of $m=-2$, too,
can have appreciable amplitude near the stellar surface.

\begin{figure}
\resizebox{0.33\columnwidth}{!}{
\includegraphics{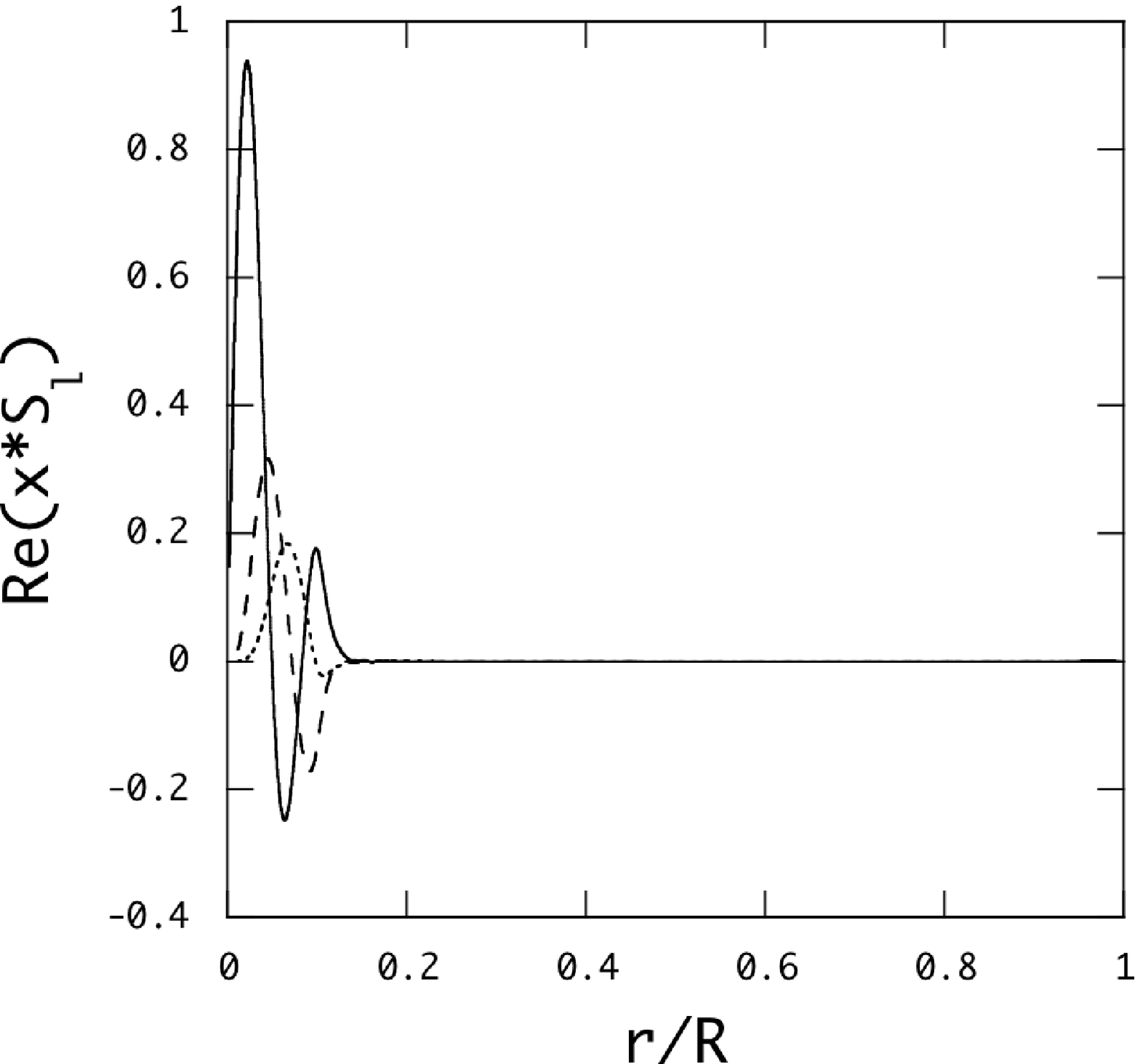}}
\resizebox{0.33\columnwidth}{!}{
\includegraphics{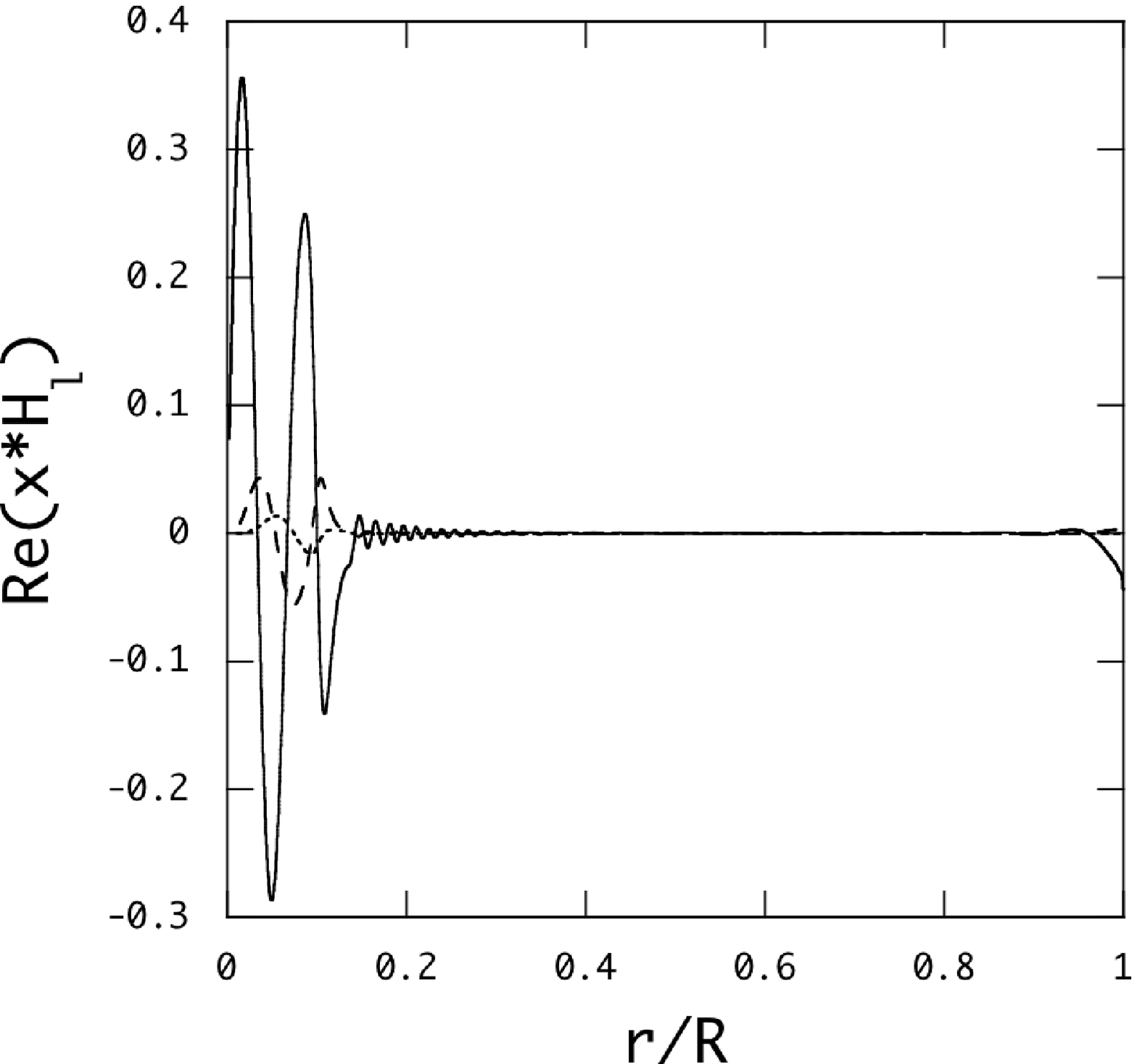}}
\caption{Same as Fig. \ref{fig:slhl_m2md1b1p2mm2_025} but 
for the $m=-2$ $B_2$ mode 
having $\overline\omega_c=(-4.24\times10^{-4}, -4.14\times 10^{-3})$ at $\overline\Omega_s=0.25$ 
(indicated by filled red squares in the middle panels of Fig. \ref{fig:m2md1b1p2}).
This mode is retrograde in the co-rotating frame of the convective core but due to the differential rotation
it couples with a prograde $g$-mode in the envelope.
}
\label{fig:slhl_m2md1b1p2mm2_B2_025}
\end{figure}

We find there exist unstable convective modes that have a co-rotation point as defined by $\omega_{\rm R}(r)=0$ in
the interior.
Fig. \ref{fig:slhl_m2md1b1p2mm2_B2_025} plots the first few expansion coefficients for the $m=-2$ $B_2$ mode with
$\overline\omega_c=(-4.24\times10^{-4}, -4.14\times 10^{-3})$ at $\overline\Omega_s=0.25$
(indicated by filled red squares in the middle panels of Fig. \ref{fig:m2md1b1p2}).
The negative $\overline\omega_{c{\rm R}}$ indicates that the mode is retrograde in a co-rotating frame in the convective core while it is prograde in the envelope because of the differential rotation, i.e., the mode has a
co-rotation point near the boundary of the convective core.
Since $|\overline\omega_{\rm I}(r)|=|\overline\omega_{c{\rm I}}|\gg\overline\omega_{c{\rm R}}$, the co-rotation point does not affect the eigenfunctions
significantly.
As indicated by the coefficients $xH_l$, the amplitudes of the convective mode could marginally reach to the stellar surface.

\subsubsection{Slightly Evolved Main Sequence Model}

\begin{figure}
\resizebox{0.33\columnwidth}{!}{
\includegraphics{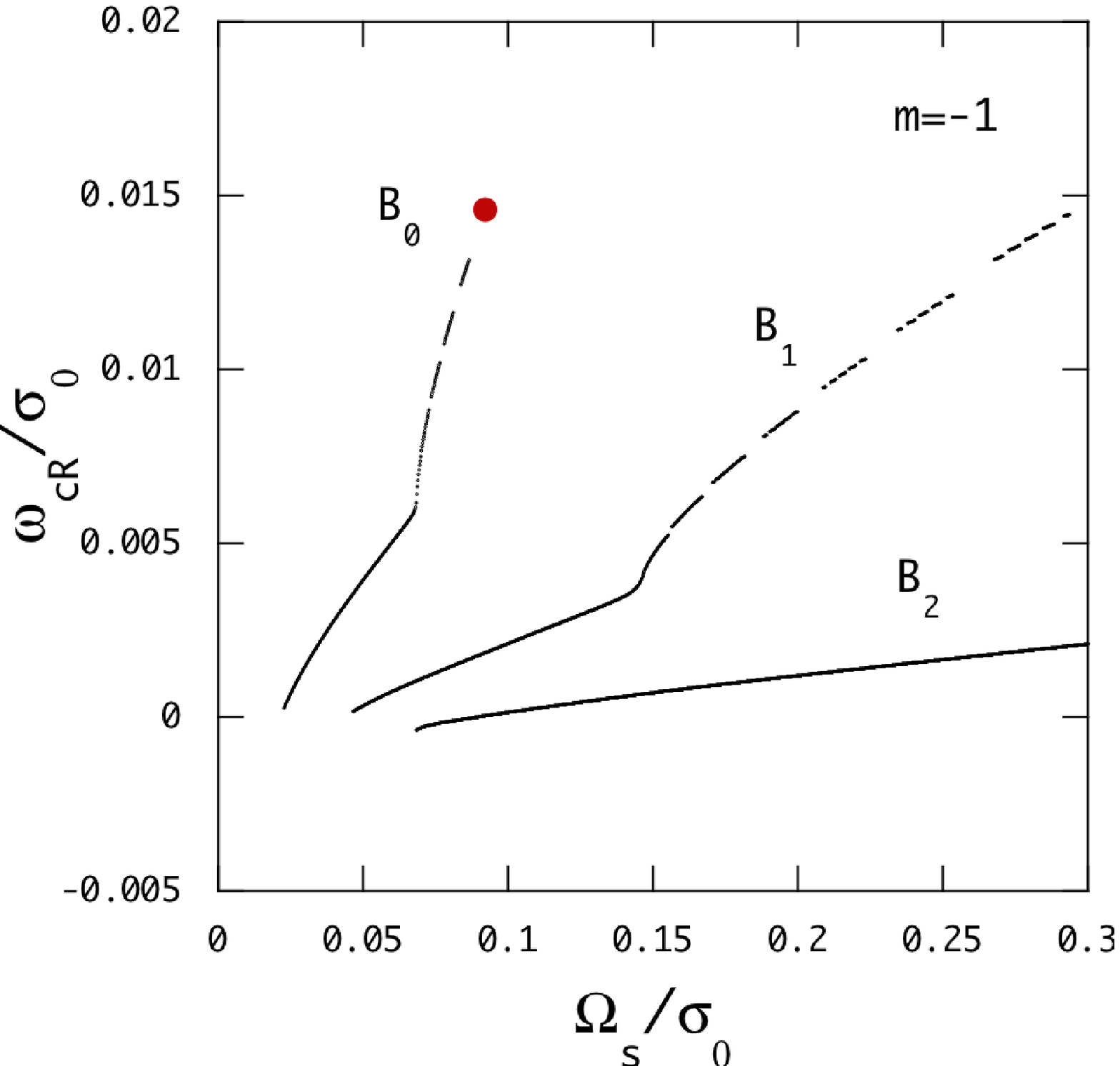}}
\resizebox{0.33\columnwidth}{!}{
\includegraphics{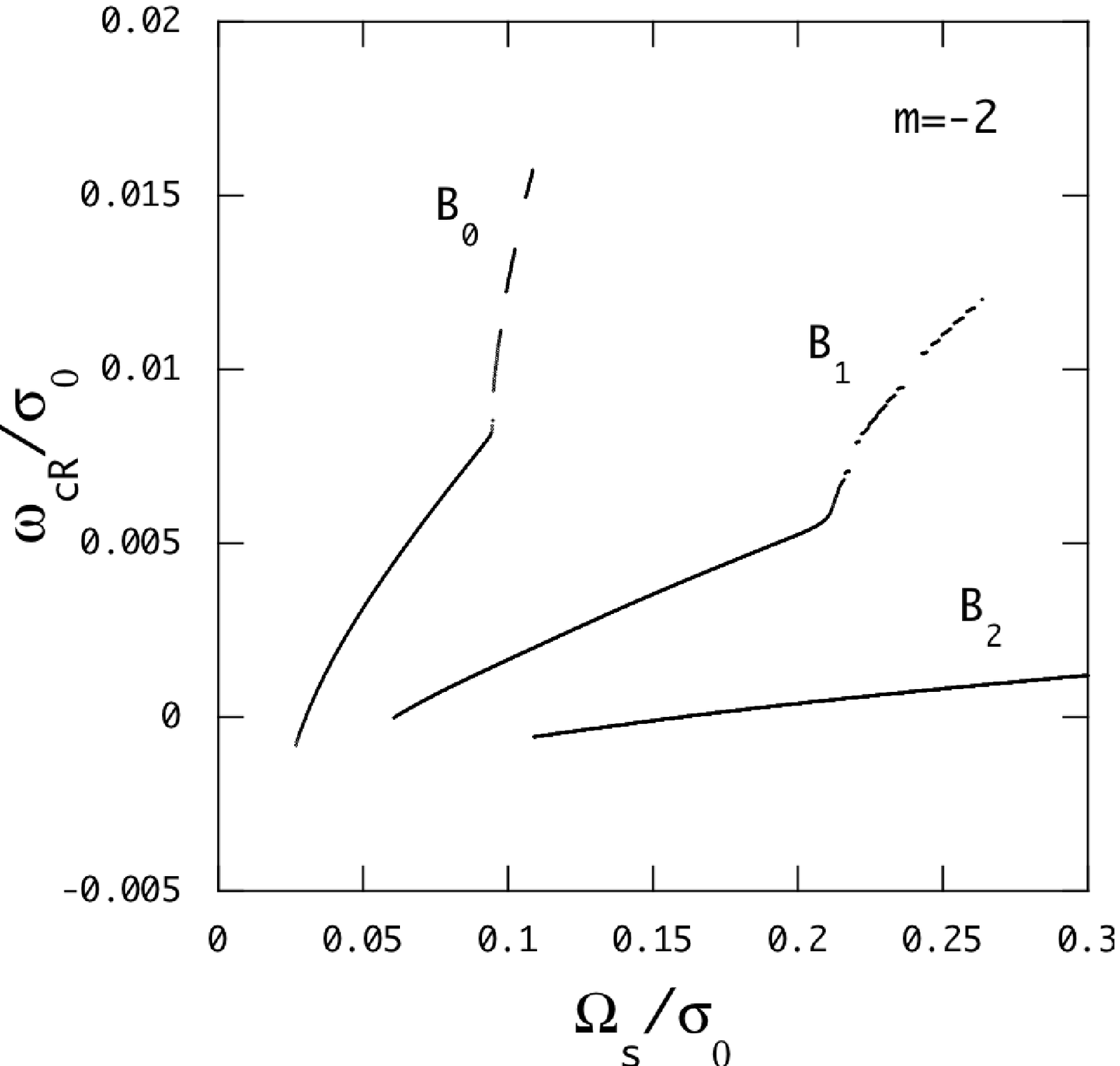}}
\resizebox{0.33\columnwidth}{!}{
\includegraphics{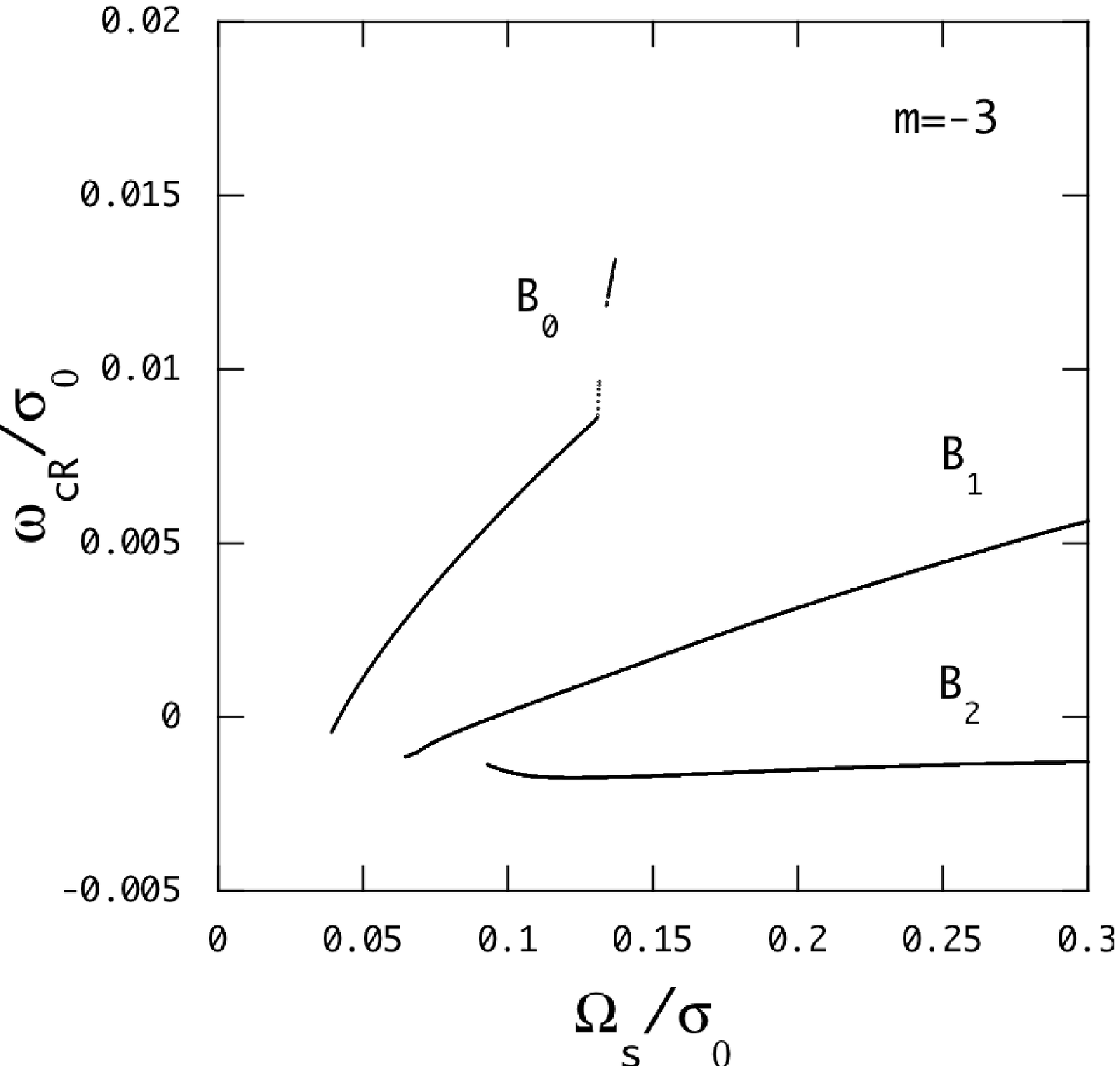}}
\resizebox{0.33\columnwidth}{!}{
\includegraphics{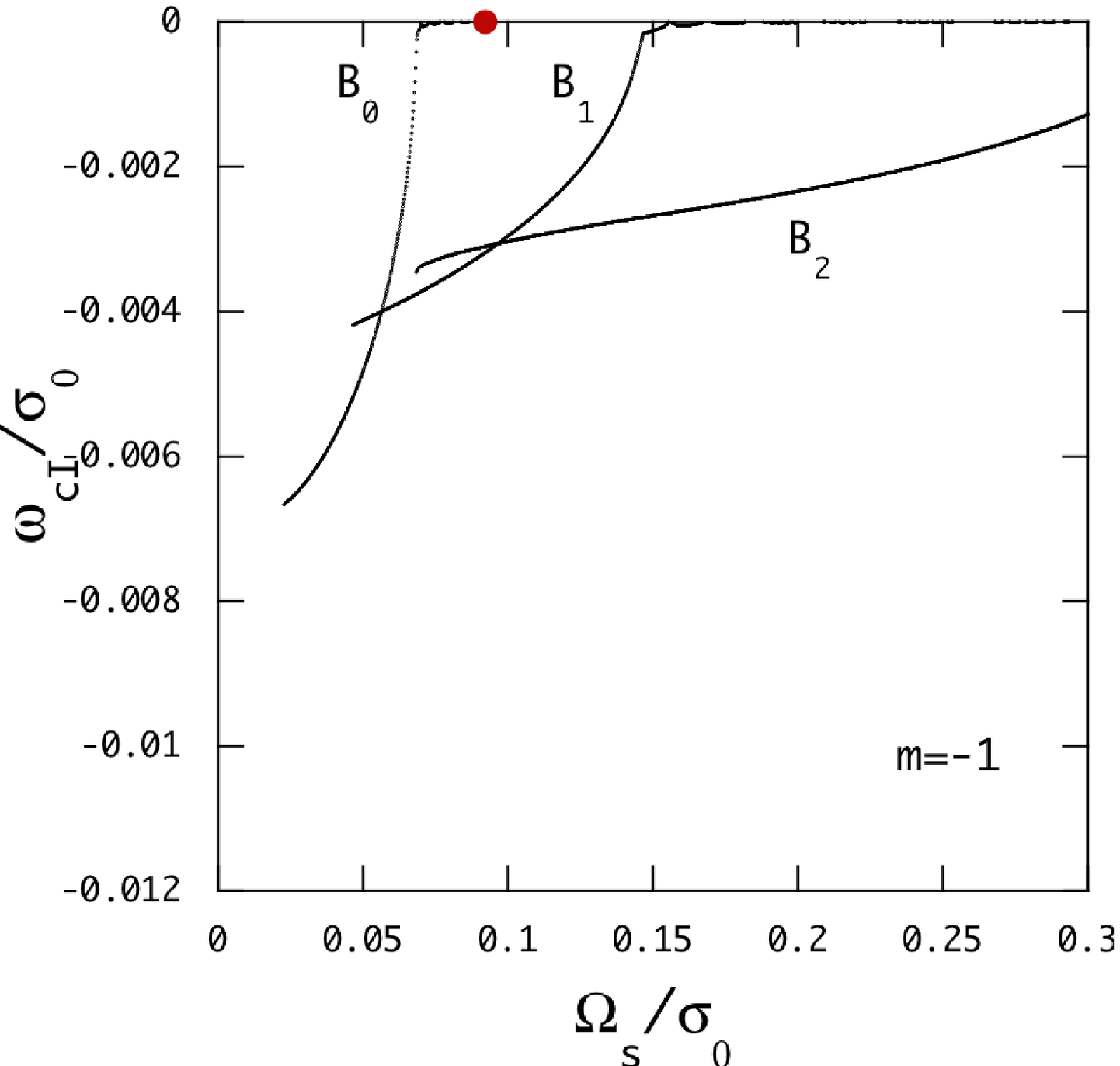}}
\resizebox{0.33\columnwidth}{!}{
\includegraphics{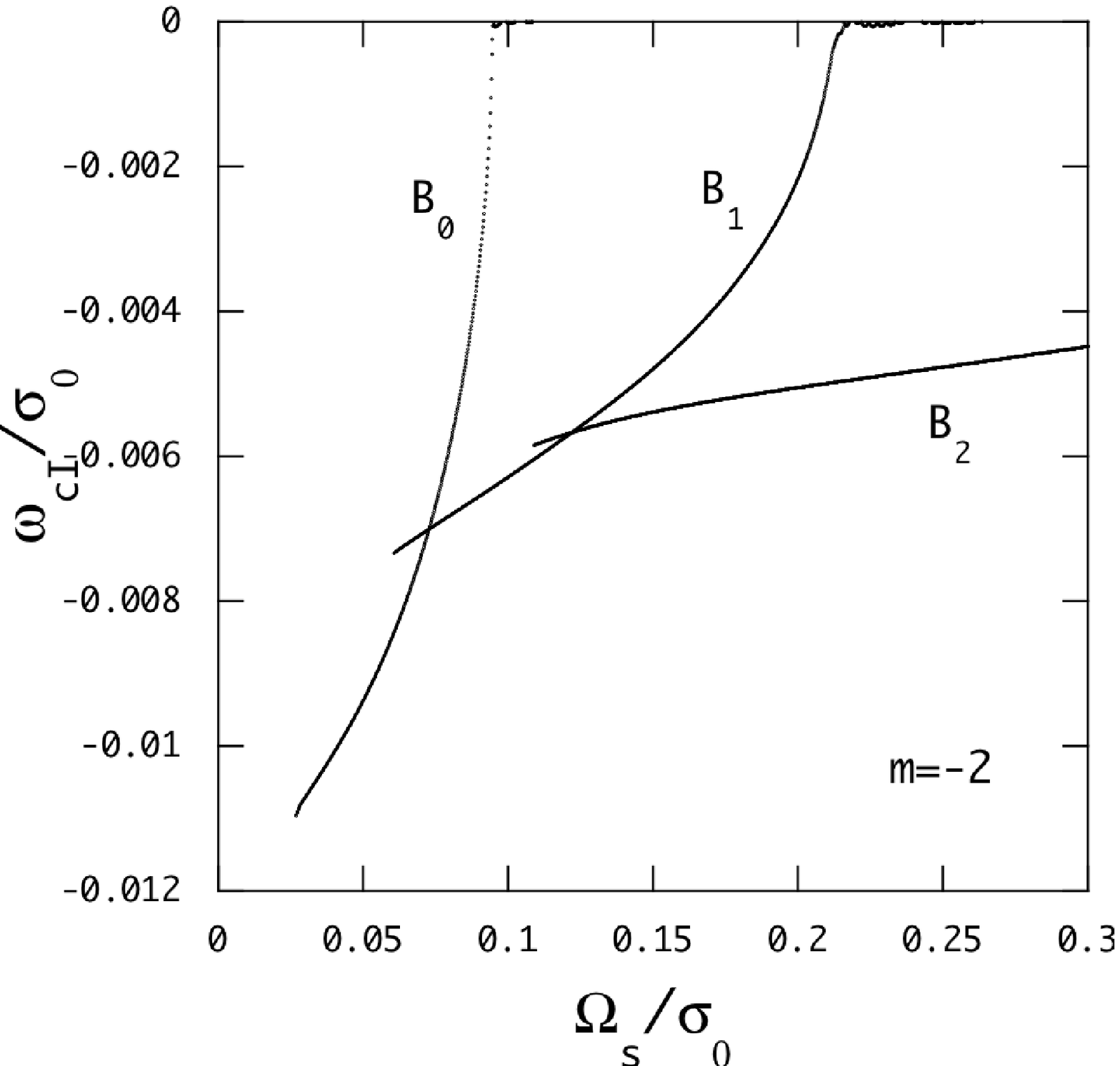}}
\resizebox{0.33\columnwidth}{!}{
\includegraphics{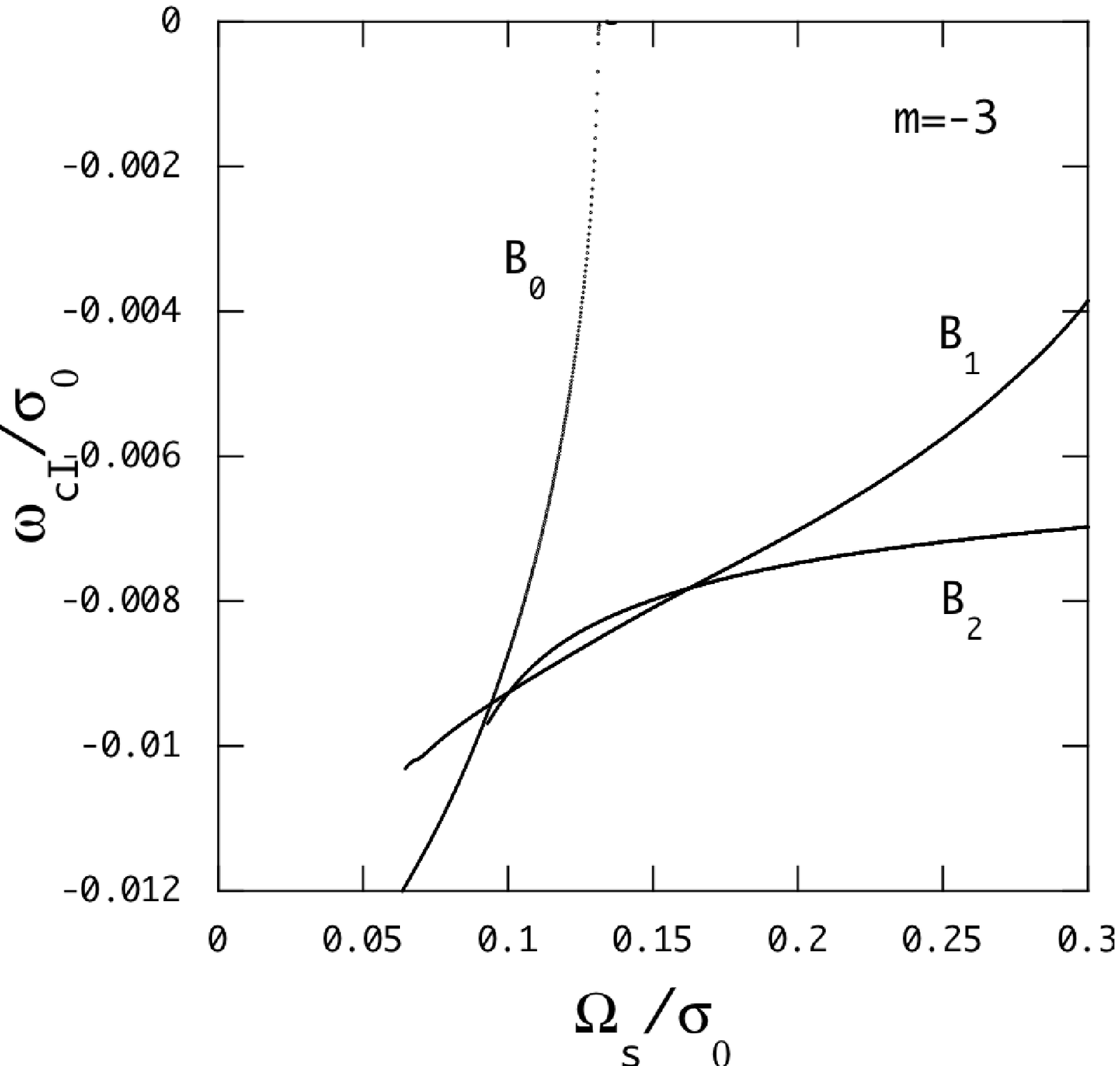}}
\caption{Same as Fig. \ref{fig:m2md1b1p2} but for a slightly evolved main sequence star with $X_c=0.5$.
Filled red circles indicate the eigenfrequency of the mode whose displacements in the interior
are shown in Fig. \ref{fig:slhl_m2md53b1p2mm1_0092}.
Narrow breaks in some lines correspond to ranges of $\Omega_s$ where modes are damped; i.e., $\omega_{c{\rm I}}>0$.
}
\label{fig:m2md53b1p2}
\end{figure}

We have also carried out non-adiabatic computation of unstable convective modes for a slightly evolved $2M_\odot$ main sequence model with $X_c=0.5$.
The convective core of the evolved model is slightly smaller compared to that of the ZAMS model and
surrounded by a $\mu$-gradient zone, as indicated by the plot of $\log_{10}\overline N^2$
in Fig. 1.

Fig. \ref{fig:m2md53b1p2} shows complex frequencies $\overline\omega_c$ of convective modes as a function of
$\overline\Omega_s$.
As shown in this figure, the behavior of $\overline\omega_{c{\rm R}}$ as a function of $\overline\Omega_s$ is very different from that found for the ZAMS model.
Instead of increasing to a peak and decreasing,
$\overline\omega_{c{\rm R}}$ keeps increasing with increasing $\overline\Omega_s$ but the increasing rate
jumps at a value of $\overline\Omega_s$ where $|\omega_{c{\rm I}}/\omega_{c{\rm R}}|$ becomes very small.
As $\overline\Omega_s$ further increases, $\overline\omega_{c{\rm R}}$
continues to increase while $-\overline\omega_{c{\rm I}}$ continues to decrease changing its sign until
the mode is completely stabilized.
For the ZAMS model there exists no indication of complete stabilization of the unstable convective modes
although we had to stop calculations when both $\overline\omega_{c{\rm R}}$ and
$|\overline\omega_{c{\rm I}}|$ become very small.
We find that as $\overline\Omega_s$ further increases after $|\overline\omega_{c{\rm I}}|$ becomes sufficiently small, 
the ratio $\omega_{c{\rm R}}/\Omega_s$
tends to be constant, which suggests that the unstable convective modes are stabilized to become
inertial modes (see below).

The gross properties of unstable convective modes in the core are quite similar between the cases of
$m=-1$, $-2$, and $-3$.
The value of $\overline\Omega_s$ at which $|\overline\omega_{c{\rm I}}|$ of
$B_0$ mode becomes sufficiently small shifts to larger values as $|m|$ increases.
We also find that the $m=-3$ $B_2$ mode have a co-rotation point.

The reason why the behavior of $\overline\omega_{c{\rm R}}$ as a function of $\overline\Omega_s$ is different
between the ZAMS model and the slightly evolved model is not clear.
Using a local analysis of waves in the short wavelength and low frequency limit, we may obtain
a dispersion relation given by (e.g., \citealt{Unnoetal1989}; see also \citealt{LeeSaio97})
\be
\omega^2\approx {N^2k_H^2+(2\pmb{\Omega}\cdot\pmb{k})^2\over k^2},
\ee
where $\omega$ is the wave frequency in the co-rotating frame, 
$\pmb{\Omega}$ is the vector of angular frequency of rotation, $\pmb{k}$ is the wave number vector with $k=|\pmb{k}|$,
and $k_H$ is the magnitude of the horizontal component of $\pmb{k}$.
As suggested by the dispersion relation, modal properties of waves in the rotating convective core
are determined by balance between buoyant effects and rotation effects. 
If buoyant effects dominate, convective modes in the core behave like oscillatory convective modes 
$\omega^2\sim \lambda N^2/k^2$ with $N^2<0$ and $\lambda<0$ and
if rotation effects dominate they behave like inertial modes with $\omega^2\sim (2\pmb{\Omega}\cdot\pmb{k})^2/k^2$, where
we have replaced $k_H^2$ by $\lambda$ assuming the traditional approximation 
\citep[e.g.,][]{LeeSaio87,LeeSaio89,LeeSaio97}.
If we used $\epsilon=10^{-3}$ in the core as in \citet{LeeSaio86}, buoyant effects represented by $N^2$ would always dominate rotation effects
$(2\pmb{\Omega}\cdot\pmb{k})^2$ so that the relation $\omega^2\approx \lambda N^2/k^2$ would describe well the wave properties
in the core.
For $\epsilon=10^{-5}$, however, the balance can be marginal and the stabilized convective modes behave like
inertial modes in the slightly evolved model.

\begin{figure}
\resizebox{0.33\columnwidth}{!}{
\includegraphics{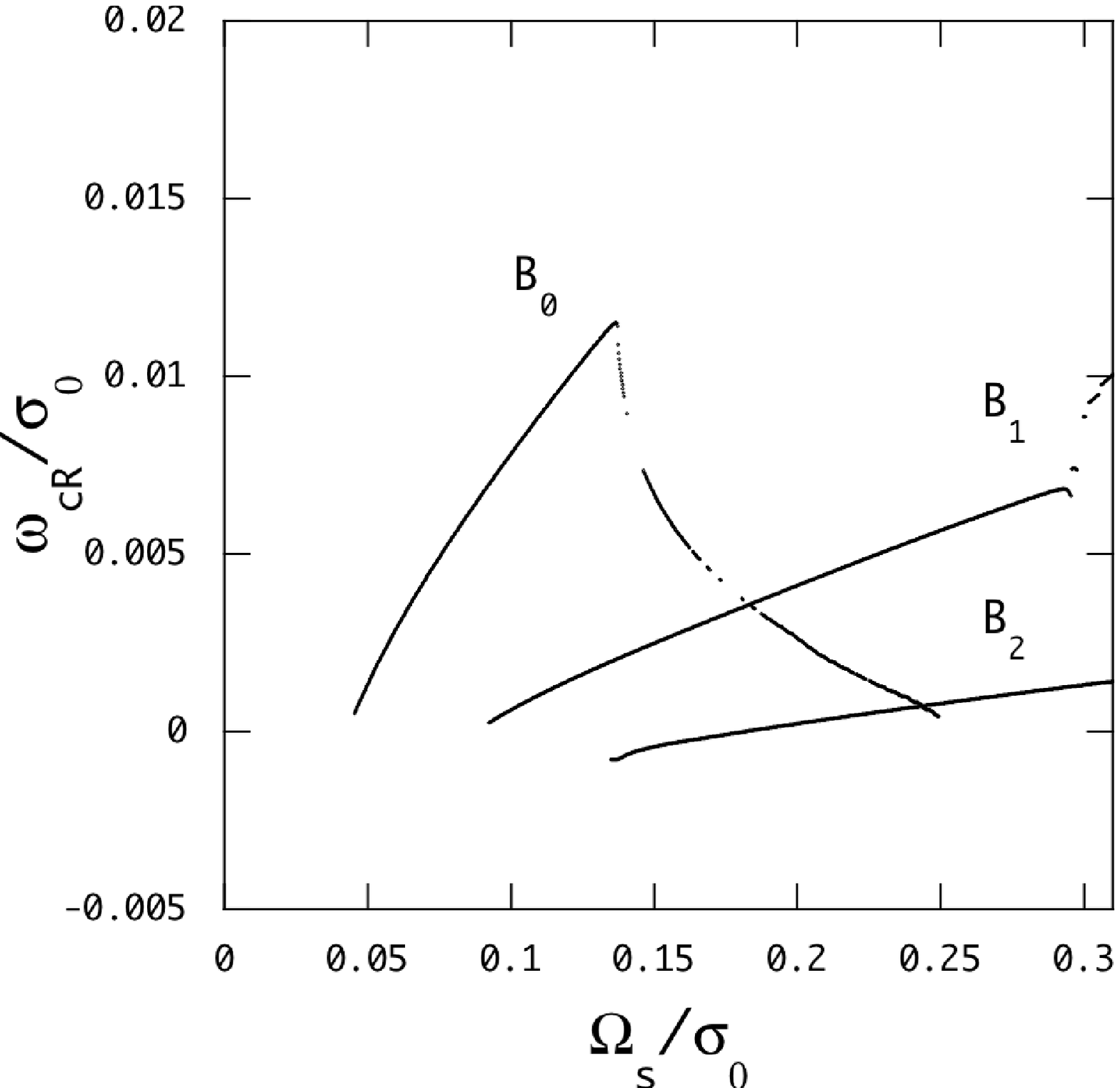}}
\resizebox{0.33\columnwidth}{!}{
\includegraphics{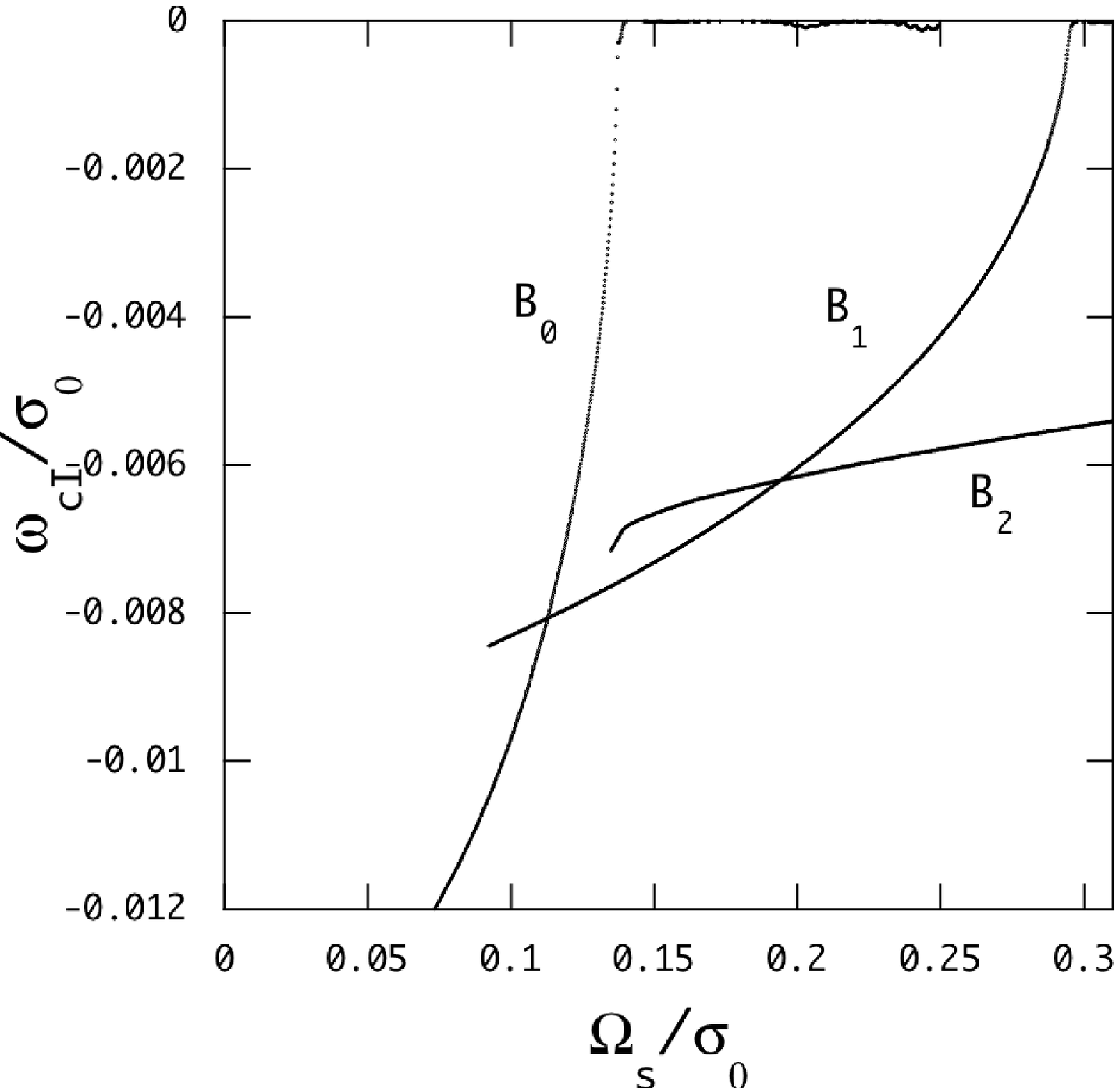}}
\caption{The same modes as those shown in the left panels for $m=-1$ of Fig.\ref{fig:m2md53b1p2} but for eigenfrequencies obtained
by assuming a larger $\epsilon=4\times10^{-5}$.
Left and right panels present real and imaginary parts, respectively.
}
\label{fig:omega_m2md53b1p2dm45mm1}
\end{figure}

To examine this interpretation,
we have computed unstable convective modes of the evolved model assuming $\epsilon=4\times10^{-5}$,
instead of $\epsilon=10^{-5}$.
We present the results for $m=-1$ in Fig. \ref{fig:omega_m2md53b1p2dm45mm1}.
It is interesting to note that $\overline\omega_{c{\rm R}}$ of $B_0$ mode has a peak as
a function of $\overline\Omega_s$, while $\overline\omega_{c{\rm R}}$ of $B_1$ mode continues to increase.
The difference may be explained as follows: Since mode $B_0$ is stabilized at a smaller value of $\overline\Omega_s$
compared with mode $B_1$, buoyant force is still significant for the former case while the
rotation effect dominates for the latter.

\begin{figure}
\resizebox{0.33\columnwidth}{!}{
\includegraphics{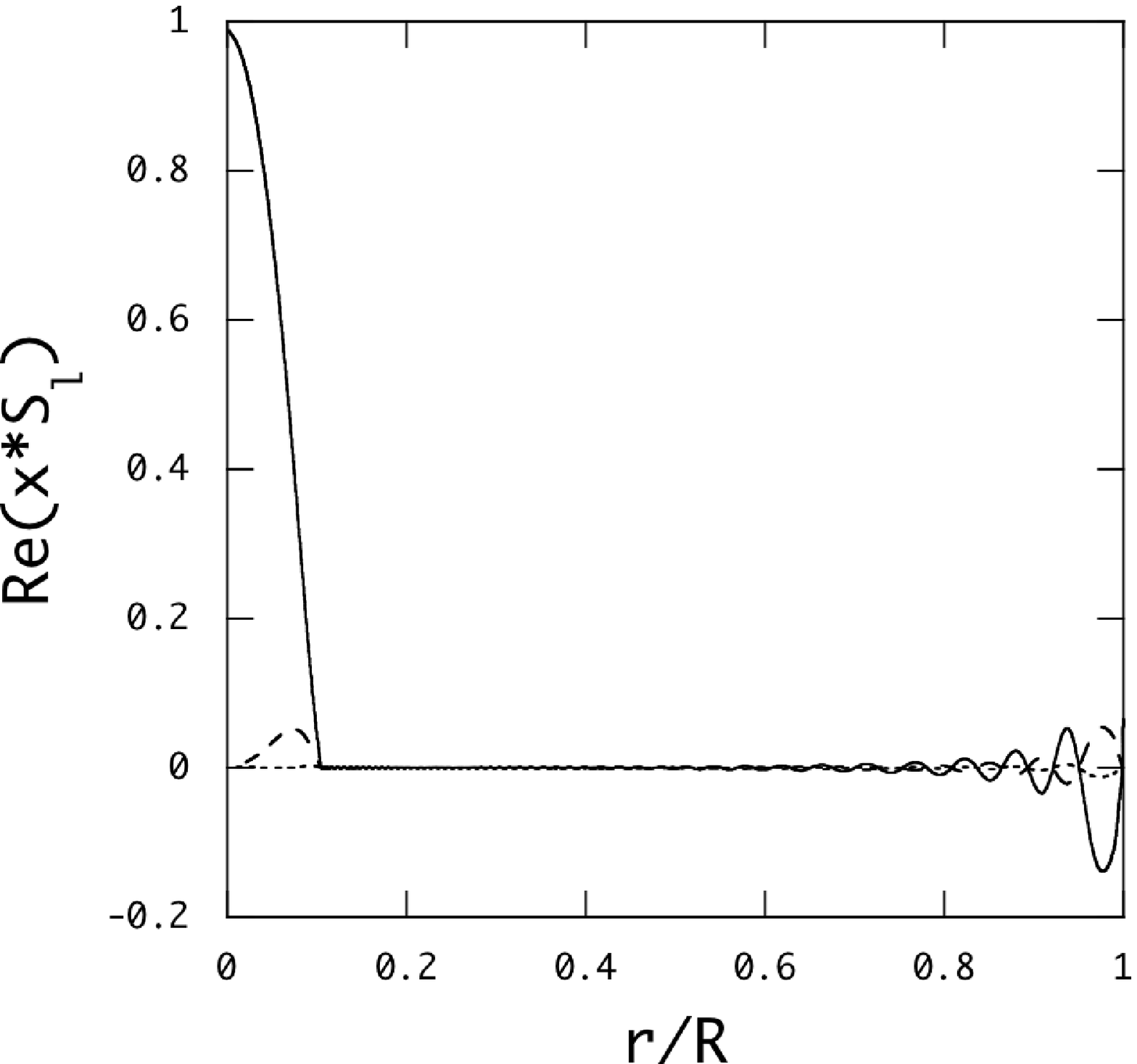}}
\resizebox{0.33\columnwidth}{!}{
\includegraphics{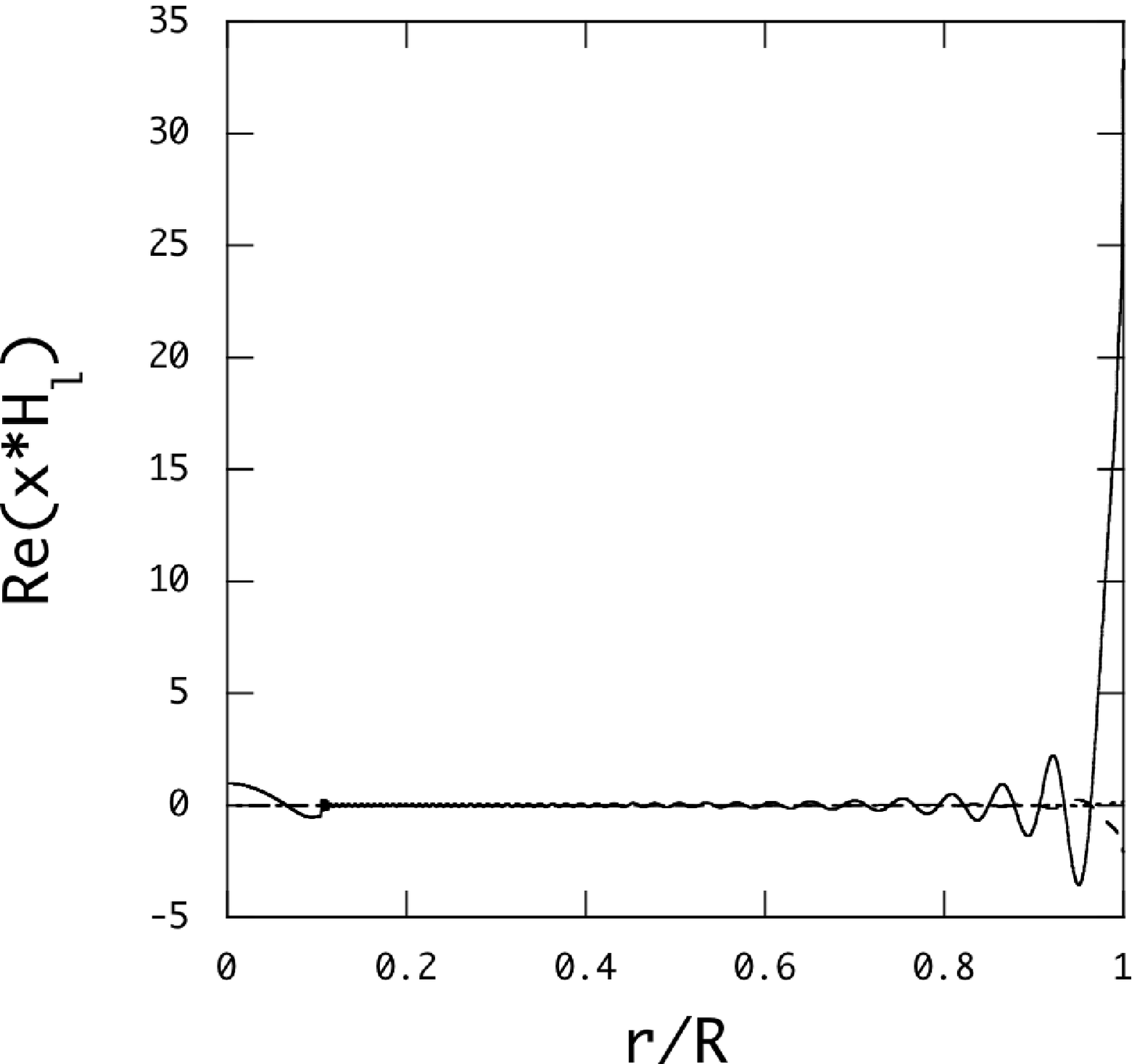}}
\caption{Real parts of the expansion coefficients for radial and horizontal displacements $xS_l$ and $xH_l$ as a function of $x=r/R$ for the $m=-1$ $B_0$ mode
with $\overline\omega_c=(1.46\times10^{-2}, -2.27\times10^{-6})$
at $\overline\Omega_s=0.092$ (indicated by filled red circles in the left panels of Fig. \ref{fig:m2md53b1p2}).
}
\label{fig:slhl_m2md53b1p2mm1_0092}
\end{figure}

For the slightly evolved model, amplitude penetration of unstable convective modes into the envelope also takes place if $|\omega_{c{\rm I}}/\omega_{c{\rm R}}|$ is sufficiently small.
Fig. \ref{fig:slhl_m2md53b1p2mm1_0092} shows an example of amplitude penetration of the $m=-1$ $B_0$ mode with $\overline\omega_c=(1.46\times10^{-2}, -2.27\times10^{-6})$ at $\overline\Omega_s=0.092$ ($\epsilon=10^{-5}$ and $b=1.2$).
The coefficient $S_{l=|m|}$ of $B_0$ mode has no nodes in the convective core.

\section{Observational Consequences}

\begin{figure}
\resizebox{0.33\columnwidth}{!}{
\includegraphics{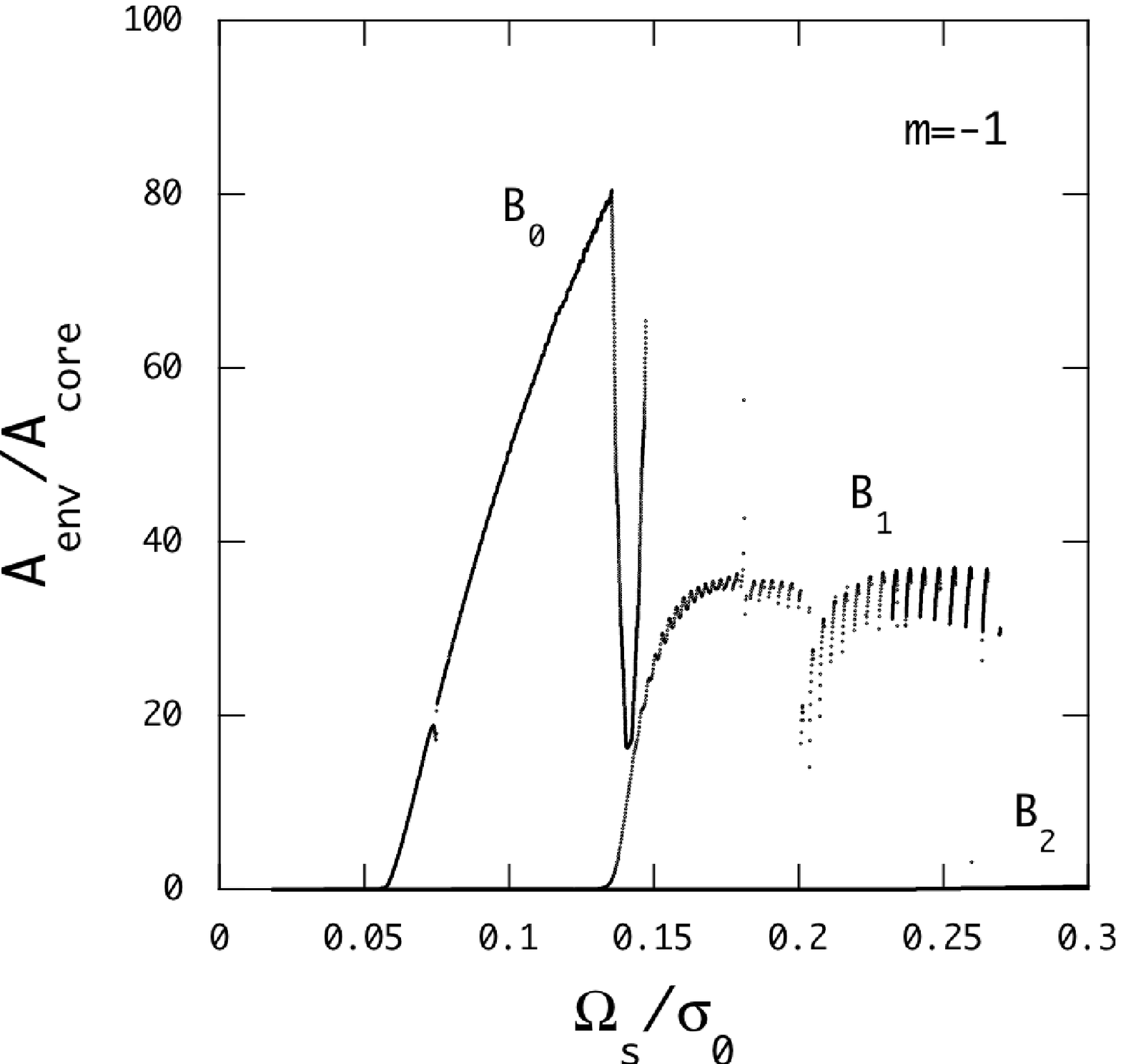}}
\resizebox{0.33\columnwidth}{!}{
\includegraphics{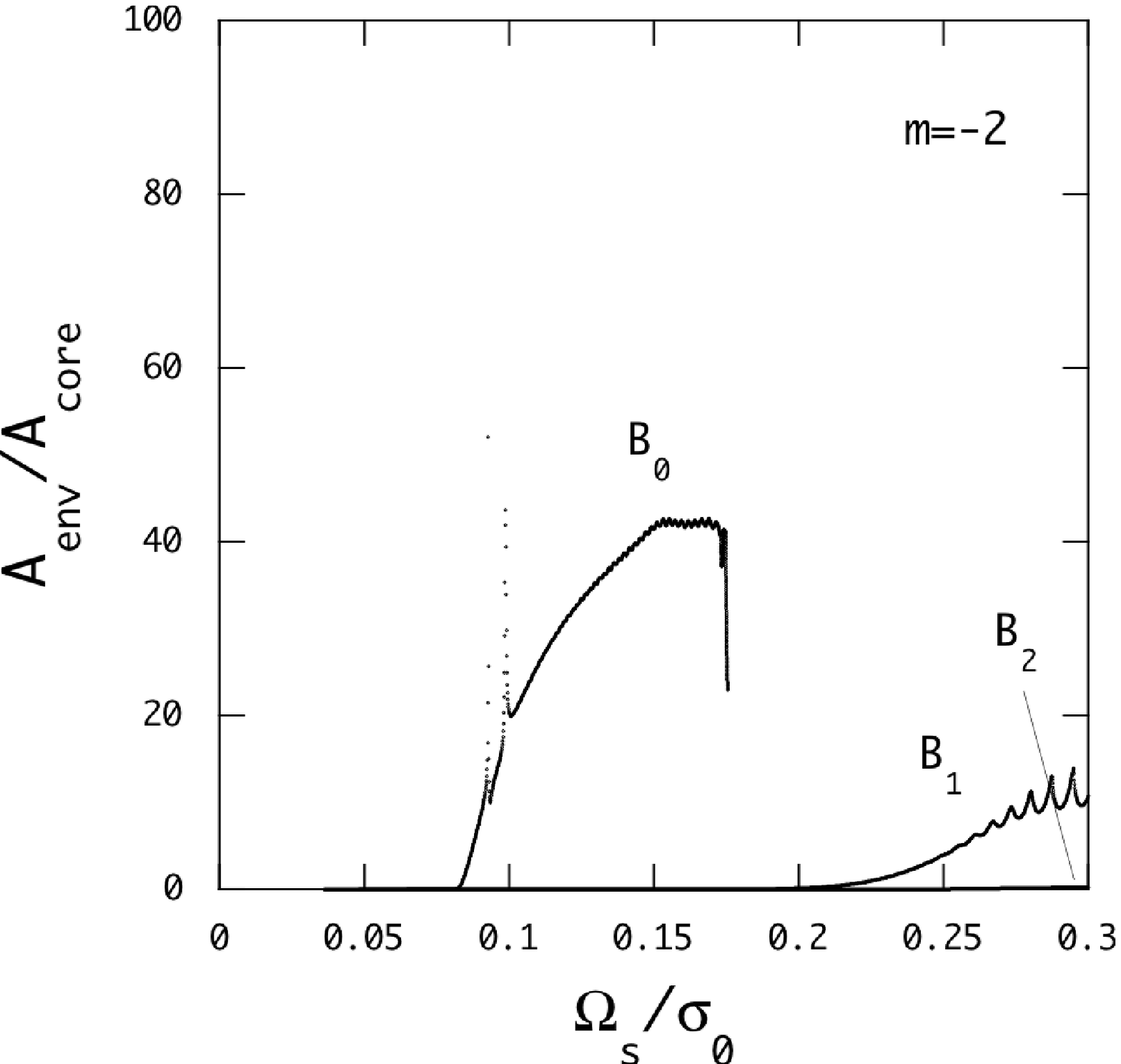}}
\resizebox{0.33\columnwidth}{!}{
\includegraphics{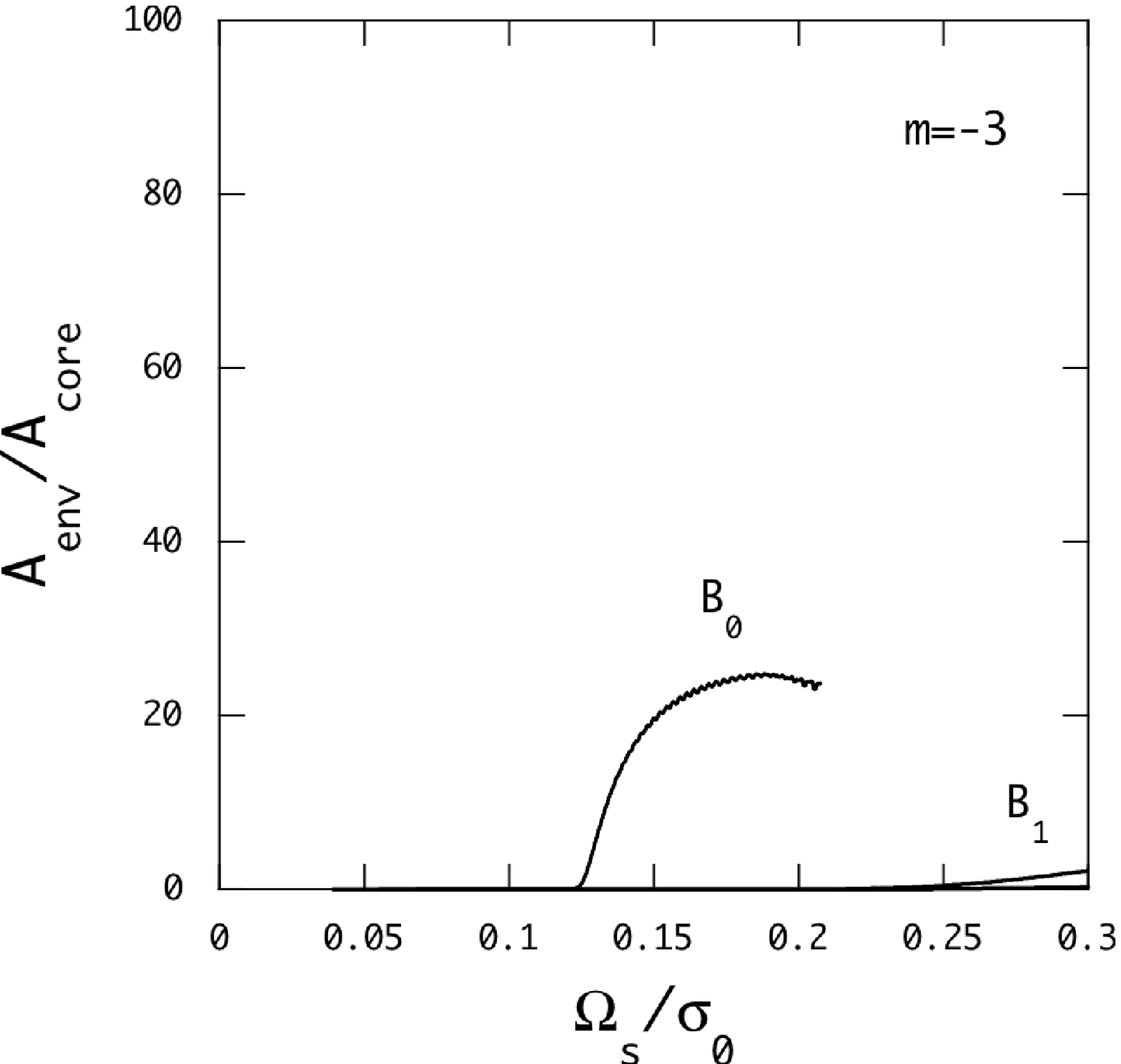}}
\caption{Envelope-core amplitude ratios $A_{\rm env}/A_{\rm core}$ against $\Omega_s/\sigma_0$ for prograde
unstable convective modes of $m=-1$, $-2$, and $-3$ in the $2M_\odot$ ZAMS model.
}
\label{fig:aeaczams}
\end{figure}

\begin{figure}
\resizebox{0.33\columnwidth}{!}{
\includegraphics{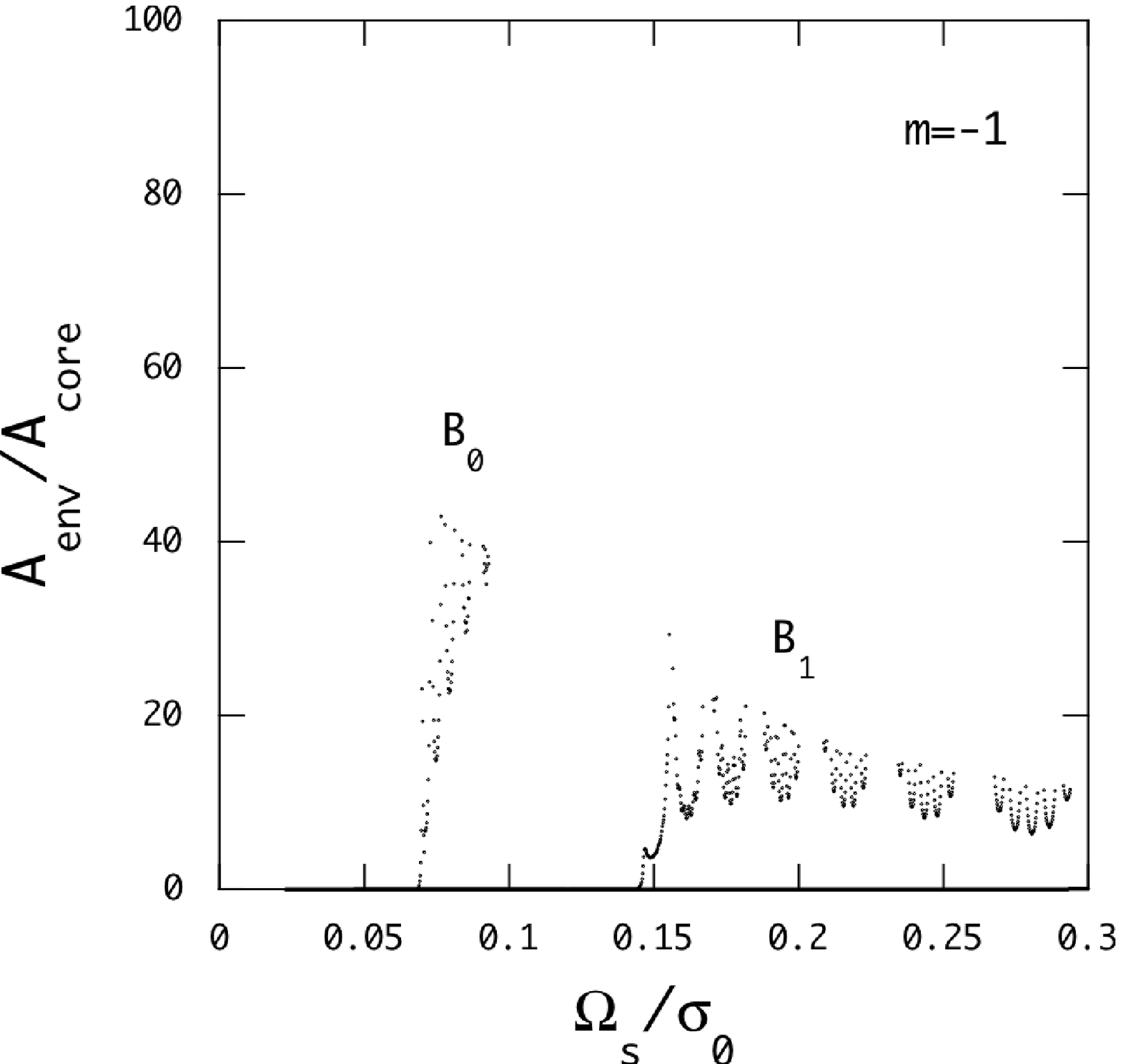}}
\resizebox{0.33\columnwidth}{!}{
\includegraphics{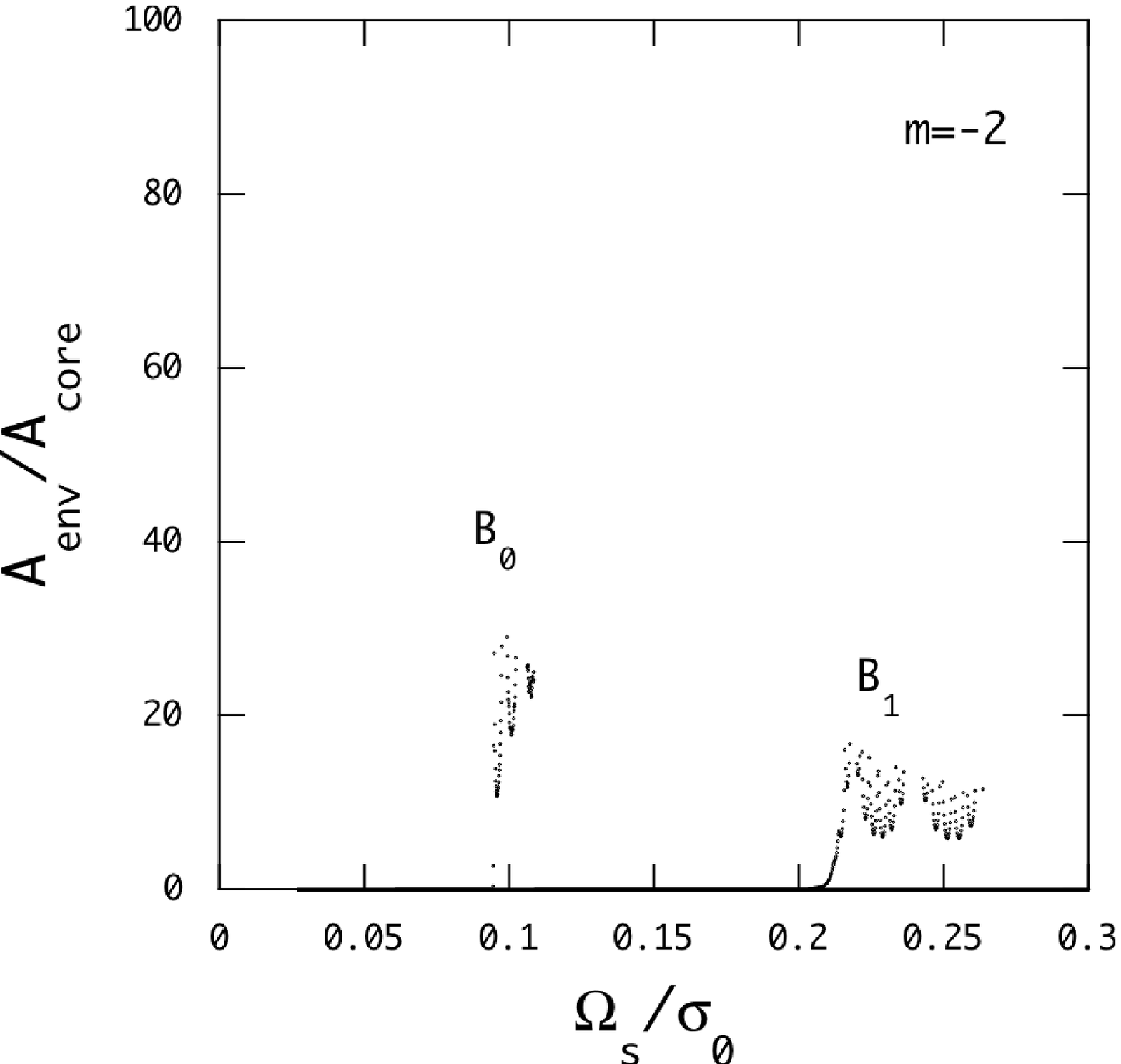}}
\resizebox{0.33\columnwidth}{!}{
\includegraphics{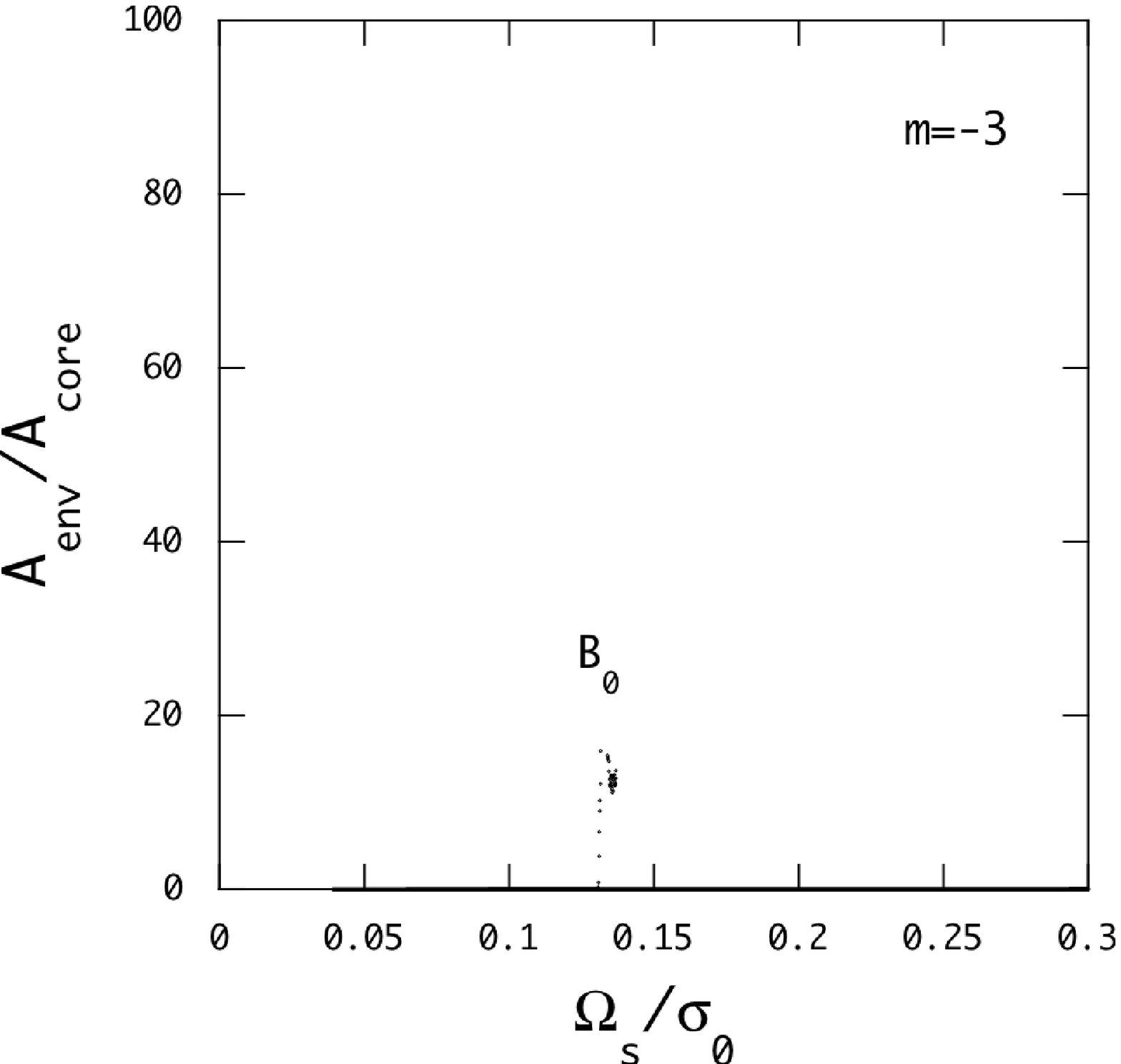}}
\caption{Same as Fig.\ref{fig:aeaczams} but for the slightly evolved main sequence model with $X_c=0.5$.
Breaks of sequences indicate the ranges of $\Omega_s$ where convective modes are damped; i.e.,
$\omega_{\rm I}>0$.
}
\label{fig:aeacmd53}
\end{figure}

In order to see when convective modes have large amplitudes in the envelope,
we have computed the ratio of 
amplitudes $A_{\rm env}$ to $A_{\rm core}$ as a function of
$\overline{\Omega}_s$ where $A_{\rm env}$ is the maximum amplitude of $|xH_{l_j}|$ in the envelope
and $A_{\rm core}$ is that of $|xS_{l_j}|$ in the core for $j=1,\cdots,~j_{\rm max}$.
The results are shown in Fig.\ref{fig:aeaczams} for the ZAMS model and in Fig.\ref{fig:aeacmd53} 
for the evolved model.
Comparing these figures with Figs. \ref{fig:m2md1b1p2} and \ref{fig:m2md53b1p2}, we find that
amplitude penetration of convective modes into the envelope, that is, $A_{\rm env}/A_{\rm core}\gtrsim 1$
takes place when $|\overline\omega_{s{\rm I}}|$ is much smaller than $|\overline\omega_{s{\rm R}}|$
and $|\overline\omega_{c{\rm R}}|$.
Note that the maximum amplitude of $|xH_{l_j}|$ occurs at the surface when $A_{\rm env}/A_{\rm core}\gtrsim1$.
The ratio $A_{\rm env}/A_{\rm core}$ for $B_0$ mode, for example, is smaller for larger $|m|$ because 
amplitude confinement into the core is stronger for larger $|m|$.
This means that prograde dipole modes with $m=-1$ will be most visible.
This property is, in fact, more pronounced, because $m=-1$ prograde sectoral mode should be least affected by geometrical cancellation on the stellar surface.
The ratios for the evolved model are in general smaller than those for the ZAMS model.
For both main sequence models, the ratios for $B_0$ and $B_1$ modes fluctuate with small amplitudes
as a function of $\overline\Omega_s$.
These fluctuations are a manifestation of resonances between convective modes and envelope $g$-modes.

As the star evolves from ZAMS, as suggested by Figs. 12 and 13, 
g-mode excitation by overstable convective modes becomes less efficient in the sense that 
the amplitude ratios $A_{\rm env}/A_{\rm core}$ become smaller and more rapid
rotation rates $\Omega_s/\sigma_0$ are required for the excitation to take place.
Our preliminary computations for the case of $2M_\odot$ model with $X_c=0.2$ confirmed that 
the rotation rates $\Omega_s/\sigma_0\gtrsim0.15$ are needed for the $B_0$ mode to excite envelope $g$-modes of $m=-1$,
while the computations also suggested that the amplitude ratios $A_{\rm env}/A_{\rm core}$ for the model of $X_c=0.2$ are 
comparable with those for the model of $X_c=0.5$.
As the main sequence star evolves, the convective core shrinks, but the central density increases and hence
the value of the integral $\int_0^{r_c}|N|d\ln r$ in the convective core does not necessarily quickly decrease with evolution.
Therefore, we expect that high-order g modes can be resonantly excited and cause rotational light modulations through to the end of main-sequence stage, being consistent with Fig.2 of \citet{Balonaetal19}.

\begin{figure}
\resizebox{0.45\columnwidth}{!}{
\includegraphics{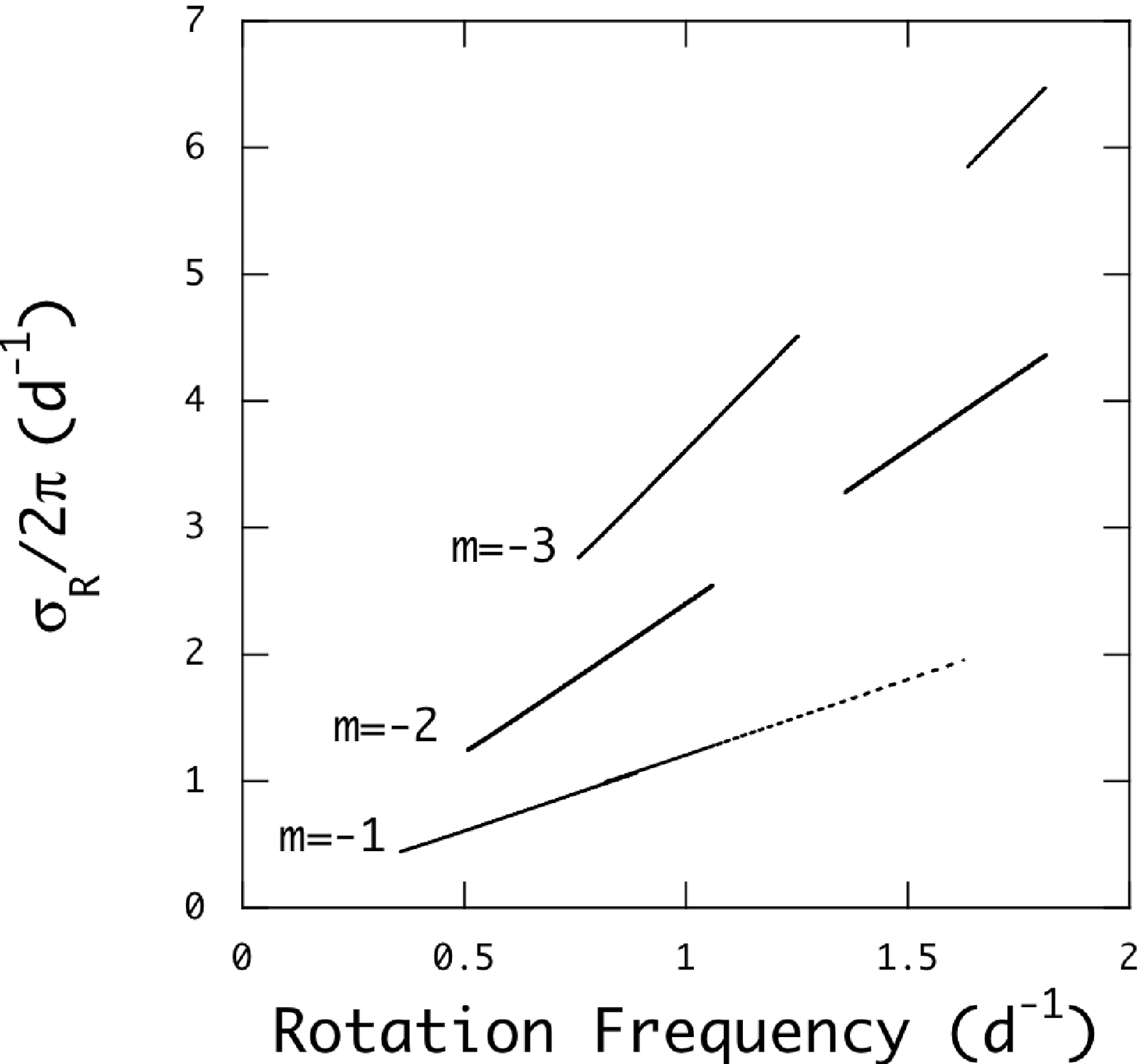}}
\resizebox{0.45\columnwidth}{!}{
\includegraphics{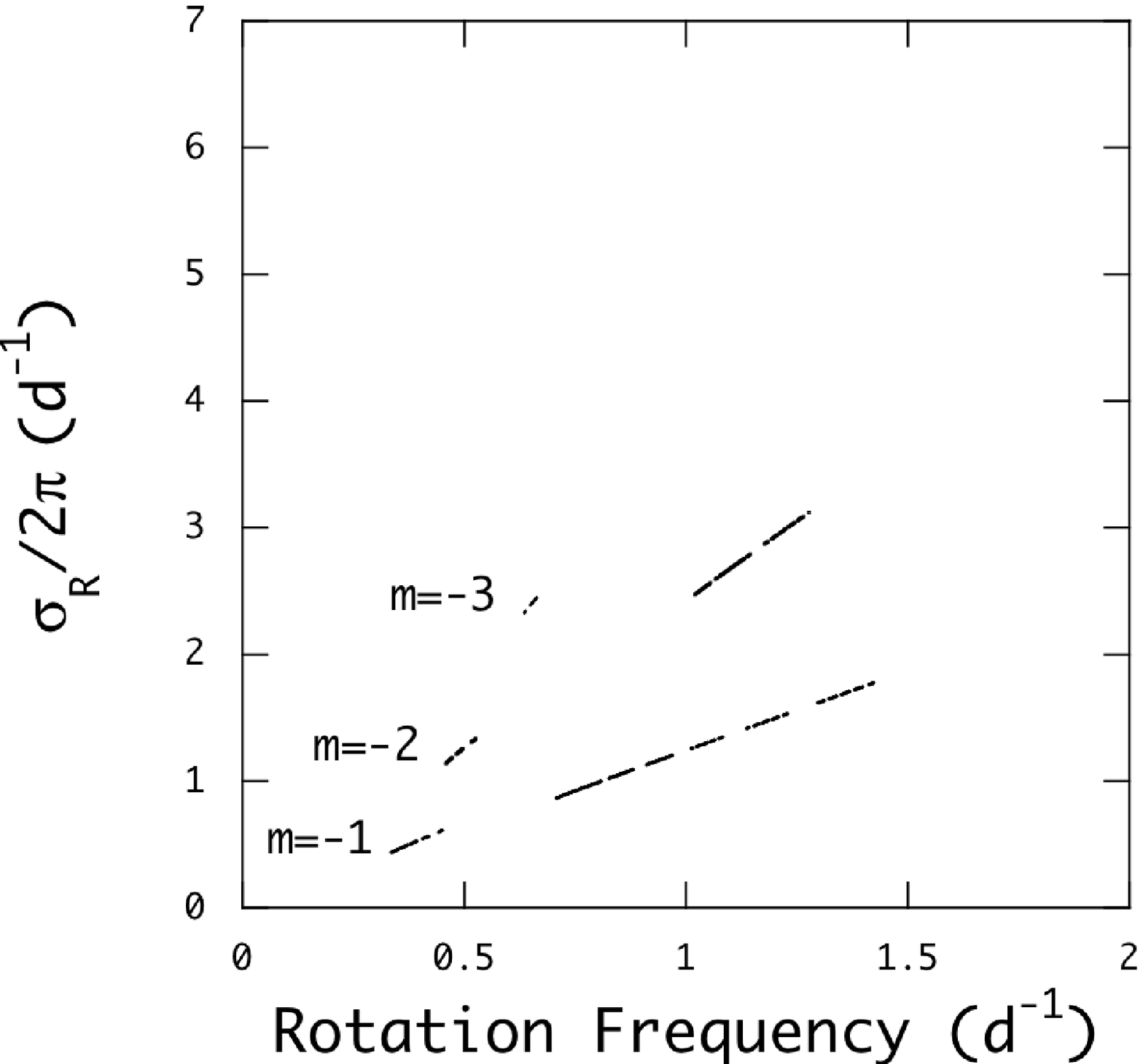}}
\caption{Cyclic frequency in the inertial frame $\sigma_{\rm R}/2\pi$ of envelope $g$-modes excited by core convective modes 
for $m=-1$, $-2$, and $-3$ 
versus surface rotation frequency $\Omega_s/2\pi$ for the $2M_\odot$ models with $X_c=0.7$ (left panel) and
$X_c=0.5$ (right panel) where only unstable convective modes
with $A_{\rm env}/A_{\rm core}\ge 1$ are plotted for $\overline\Omega_s\le 0.3$.
A weak differential rotation of $b=1.2$ is assumed.
}
\label{fig:sigma-omega}
\end{figure}

In Fig. \ref{fig:sigma-omega}, we plot cyclic frequency $\sigma_{\rm R}/2\pi$, in the inertial frame, of unstable convective modes
that are likely observable, satisfying the condition $A_{\rm env}/A_{\rm core}\ge 1$.
The left and right panels of this figure are for the ZAMS model and the evolved model, respectively.
Since the frequency $\omega_c$ in the convective core is 
much smaller than $|m\Omega_s|$, we have $\sigma_{\rm R}\sim |m\Omega_c|$ in the inertial frame
for all convective modes having large amplitudes in the envelope.
Note that for the ZAMS model the wide gaps along the loci of $m=-2$ and $-3$ convective modes are due to
the fact that we had to stop calculating convective modes when
both $\overline\omega_{c{\rm R}}$ and $|\overline\omega_{c{\rm I}}|$ become very small.
On the other hand, the wide gaps for the evolved model appear because 
the convective modes are completely stabilized (i.e., damped oscillation with $\omega_{\rm I}>0$) when $\overline\Omega_s$ gets sufficiently large.
Besides the wide gaps there appear many narrow gaps along the locus, for example,
of the $m=-1$ $B_1$ mode for the ZAMS model.
These narrow gaps are caused by frequent stability changes 
with increasing $\Omega_s$.
This kind of narrow gaps are also found for the evolved model.
Fig. \ref{fig:sigma-omega} indicates that we would expect rotational modulations and their harmonics in most early type
main-sequence stars, corresponding $|m|$ times rotation frequency of the convective core which rotates
slightly faster that the envelope.

A Fourier spectrum of the g modes excited by convective modes would show a few peaks associated with low $|m|$ g-modes 
and the number of the peaks depend on
the rotation frequency $\Omega_s$.
When $\Omega_s$ is small, we will find only a single peak produced by
an $m=-1$ g-mode although there exist a limiting value of $\Omega_s$, above which
the $m=-1$ g-mode will be observable.
As the rotation frequency $\Omega_s$ increases, the number of peaks will increase to two or more than two, depending
on $\Omega_s$.
In most cases, we will have two peaks associated with g mode frequencies $\sigma_{m=-1}$ and $\sigma_{m=-2}$, for which 
the ratio $\sigma_{m=-2}/\sigma_{m=-1}\approx 2$.
We expect that the amplitude at $\sigma_{m=-1}$ will be higher than
that at $\sigma_{m=-2}$ since the ratios $A_{\rm env}/A_{\rm core}$ are, in general, higher for $m=-1$ that for $m=-2$
and the effect of geometrical cancellation is smaller for the $m=-1$ mode.
As suggested by the left panel of Fig. 14, it can happen for a range of $\Omega_s$
that the two peaks are associated with g modes
of $\sigma_{m=-1}$ and $\sigma_{m=-3}$, for which the ratio $\sigma_{m=-3}/\sigma_{m=-1}\approx 3$.
However, the $m=-3$ mode would be hardly visible because the geometrical cancellation on the stellar surface should be large.

It is also interesting to note that for $m=-1$, there exist ranges of $\Omega_s$ in which
both $B_0$ and $B_1$ (or $B_1$ and $B_2$) convective modes can excite envelope g modes.
If this is the case, we will find two closely separated peaks around the frequency $\approx \Omega_c$.

According to Table 1 of \citet{Balonaetal19} the majority of B-type stars  classified as 'ROT' (rotational variable) have only one harmonics (or less) in the Fourier spectra.  This corresponds to  that mostly $m=-1$ and $-2$ sectoral g modes are observable, while g modes of $|m| > 2$ are hardly visible even if excited.

\begin{figure}
\resizebox{0.33\columnwidth}{!}{
\includegraphics{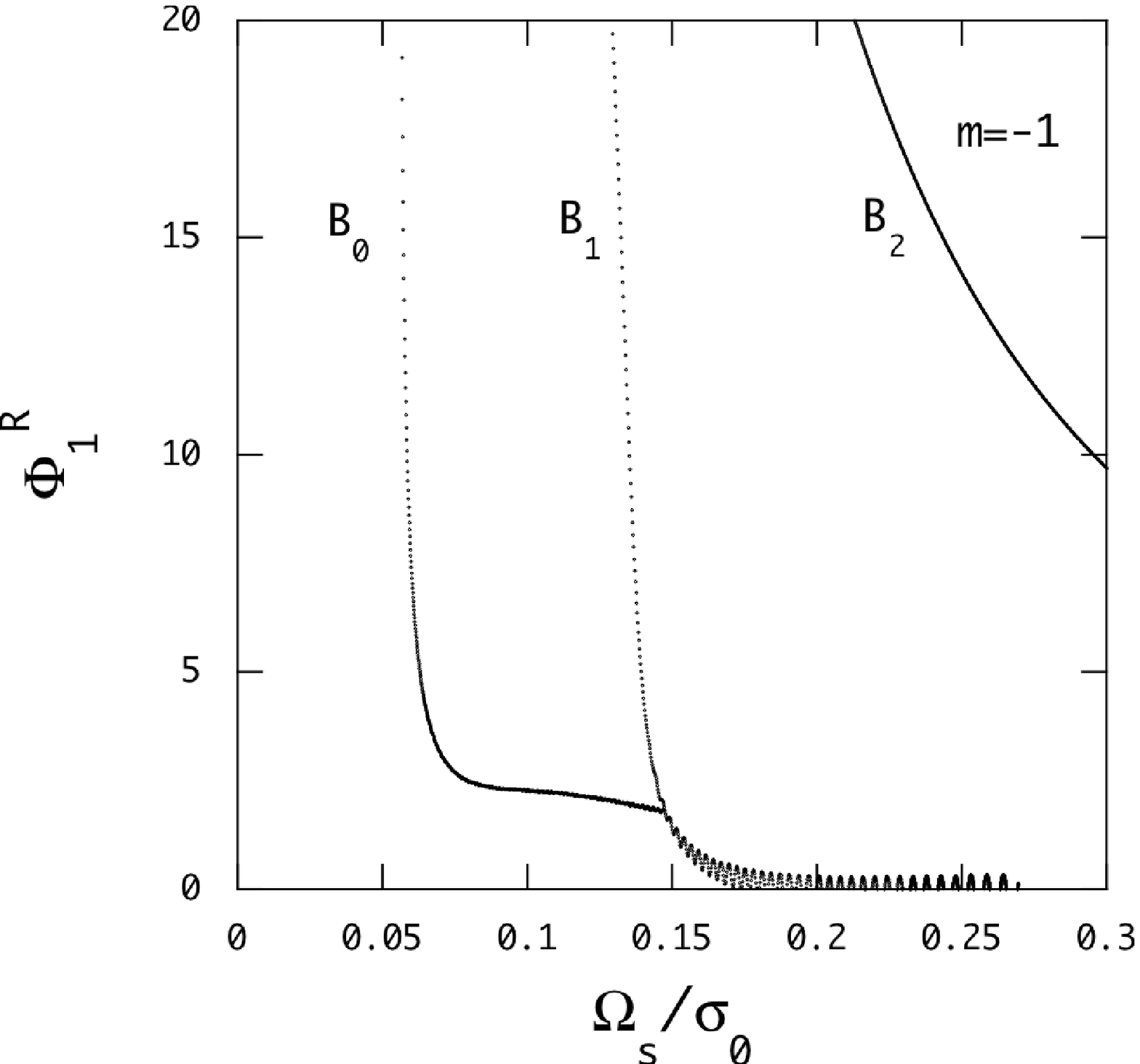}}
\resizebox{0.33\columnwidth}{!}{
\includegraphics{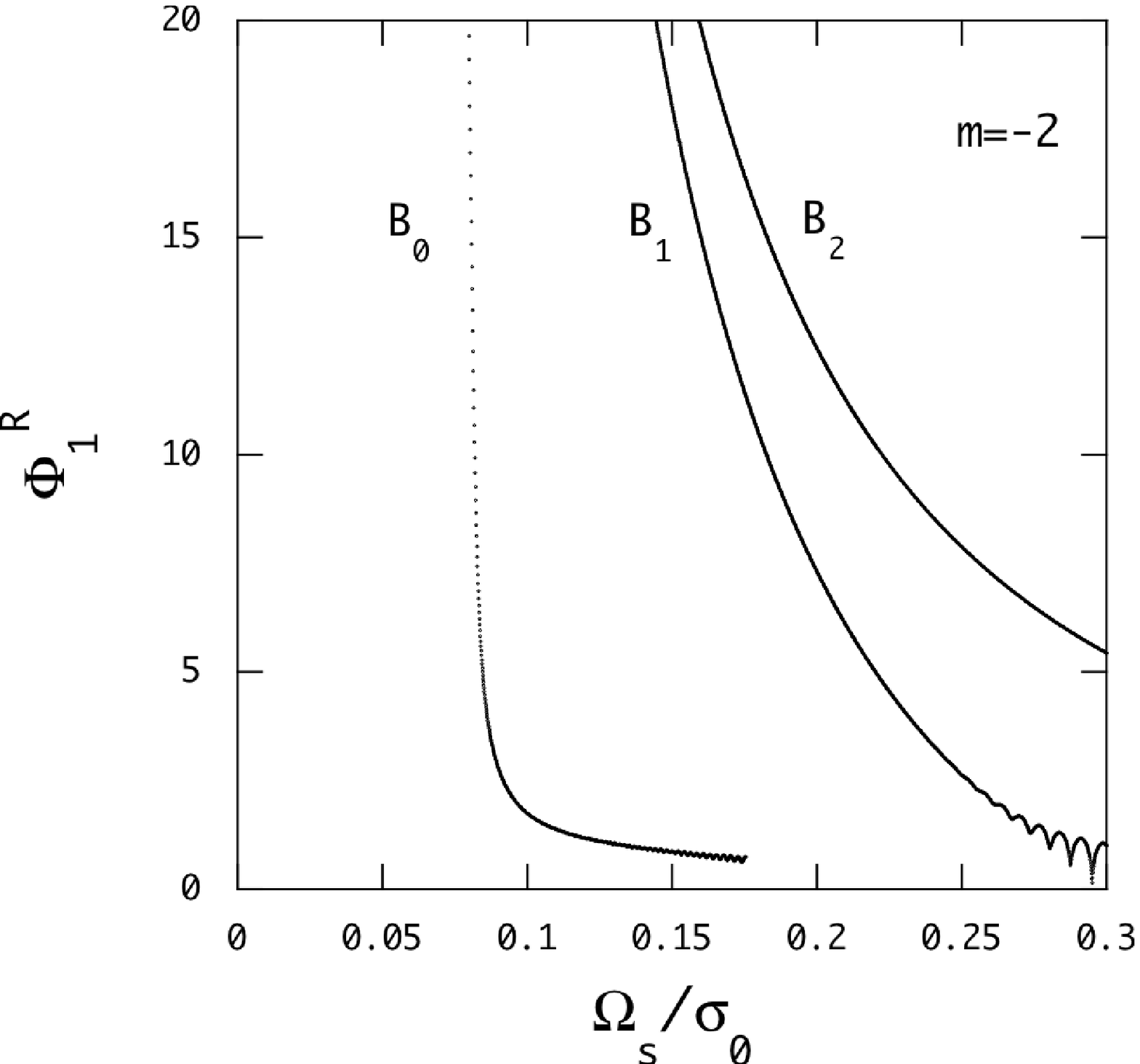}}
\resizebox{0.33\columnwidth}{!}{
\includegraphics{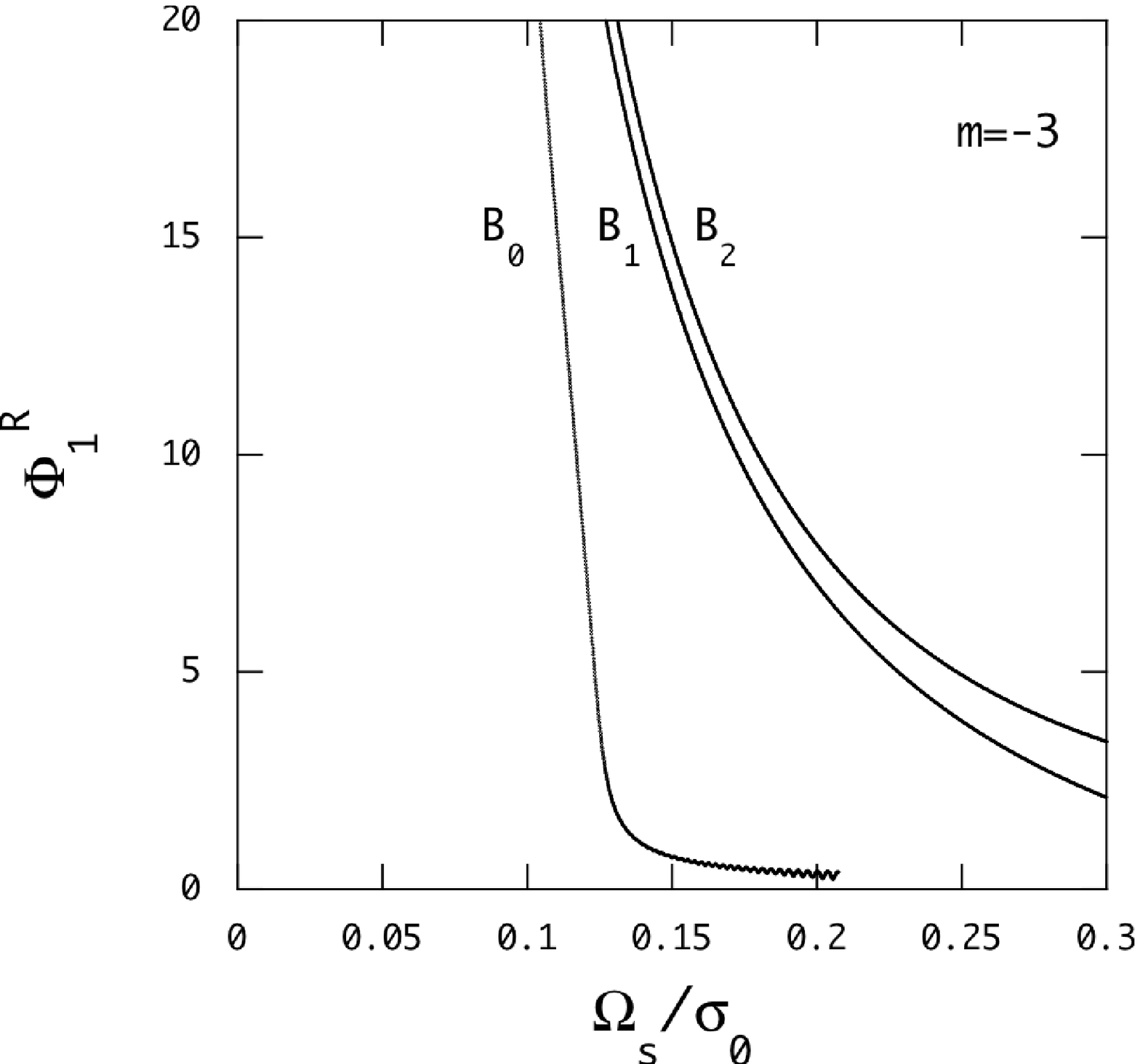}}
\caption{$\Phi_1^R$ as a function of $\Omega_s$ for low $|m|$ unstable convective modes 
in the ZAMS model for the case of the differential rotation with $b=1.2$.
}
\label{fig:phi1}
\end{figure}

Let us discuss amplitude penetration of convective modes using an asymptotic method.
We assume that the frequency $\overline\omega_c$ of convective modes is determined within the convective core and that
the convective modes with the shifted frequency $\overline\omega_s=\overline\omega_c+m(\overline\Omega_s-\overline\Omega_c)$ resonantly excite gravity waves in the envelope.
As an asymptotic treatment of non-radial oscillations suggests,  
the phase of such gravity waves having a frequency $\overline\omega_s$
in the envelope may be given by
\be
\Phi^r(\omega_s)=\int_{r_c}^rk_r(\omega_s)dr\approx
\Phi_0^r+\rmi\Phi_1^r,
\ee
where $r_c$ is the radius of the convective core and for low frequency $g$-modes in the envelope
\be
k_r(\omega)
\approx \sqrt{-{rA\over r^2}{\lambda\over c_1\overline\omega^2}}={\sqrt{\lambda}\over r}{\overline N\over\overline\omega}, \quad
\Phi_0^r=\int_{r_c}^rk_r(\omega_{s{\rm R}})dr, \quad \Phi_1^r=-{\omega_{s{\rm I}}\over\omega_{s{\rm R}}}\Phi_0^r,
\ee
and we have substituted $\lambda$, instead of $l(l+1)$, for rotating stars
under the traditional approximation \citep[e.g.,][]{LeeSaio97}.
The quantity $\int_{r_c}^Rk_rdr/\pi$ may be regarded as 
the number of radial nodes in the envelope.
Since the term $ (\omega_{s{\rm I}}/\omega_{s{\rm R}})\int_{r_c}^rk_rdr$
describes decays of wave amplitudes with wave propagation in the envelope, we may assume that gravity waves excited by
core convective modes can reach to the stellar surface if 
\be
\left|\Phi_1^R\right|
\lesssim 1.
\label{eq:phi1}
\ee
Note that $\omega_{s{\rm I}}=\omega_{c{\rm I}}$ but $\omega_{s{\rm R}}$ is larger than $\omega_{c{\rm R}}$
if the core rotates faster than the envelope.
Fig. \ref{fig:phi1} plots $\Phi_1^R$ as a function of $\overline\Omega_s$ for low $|m|$ unstable convective modes 
in the ZAMS model for a weak differential rotation with $b=1.2$.
As shown in the figure, at rapid rotation rates,
$\Phi_1^R$ decreases to become $\sim 1$, suggesting that amplitude penetration of the convective modes into the envelope takes place.
Comparing Figs. \ref{fig:phi1} and \ref{fig:aeaczams}, we find that predictions by the condition
$|\Phi_1^R|\lesssim 1$ for amplitude penetration are consistent with those by the condition $A_{\rm env}/A_{\rm core}\gtrsim 1$.
Note that in the case of uniform rotation with $b=1$, the minimal values of $\Phi_1^R$ are of order of $\sim10^3$, 
and no amplitude penetration into the envelope can be expected.
If we write $\Phi_1^R$ as
\be
\Phi_1^R=-{\omega_{s{\rm I}}\over\omega_{s{\rm R}}}{\sqrt{\lambda}\over\omega_{s{\rm R}}}\int_{r_c}^R N {dr\over r},
\ee
we understand from this equation that even-parity and prograde sectoral $g$-modes, for which $\lambda$ can be smallest
for a given $m$ \citep[e.g.,][]{LeeSaio97}, are most probable modes excited by unstable convective modes
to have significant amplitudes in the envelope.
This may justify our analysis which has been focused on even-parity prograde convective modes.

\section{Conclusion}

We have computed unstable convective modes in $2M_\odot$ main-sequence models, assuming uniform rotation and
a weak differential rotation, in which the convective core is rotating slightly faster than the radiative envelope.
Convective modes get very weakly unstable with very small growth rates (i.e., $|\omega_{c{\rm I}}/\omega_{c{\rm R}}|\ll 1$)
if the rotation speed is sufficiently high.
Then, in the models with the differential rotation, the amplitudes of unstable prograde convective modes in the core can penetrate into the radiative envelope 
of the stars as a result of resonance with high radial order $g$-modes in the envelope.
In other words, high radial order $g$-modes in the radiative envelope are resonantly excited by weakly unstable convective modes 
generated in the core which rotates slightly faster than the envelope.
We find such amplitude penetration of the modes to occur 
if rotation frequency is higher than about 0.3d$^{-1}$ in our $2M_\odot$ models.

If low $|m|$ $g$-modes in the envelope of rotating early-type stars are 
resonantly excited by weakly unstable convective modes in the core,
they will be observed as low frequency variabilities having frequencies
$\sigma\approx -m\Omega_c$ where $\Omega_c$ is the rotation frequency of the convective core.
Because a single frequency is associated with each $m$ (see Fig. \ref{fig:sigma-omega}), these $g$-mode variations would be observed as
a rotational modulation and its harmonics.
This property of resonantly excited $g$-mode oscillations nicely explains the rotational modulations and their harmonics
detected in many early type main sequence stars (e.g., \citealt{Balona19}; \citealt{Balonaetal19}).

For B-type stars there exists a longstanding unsolved problem, that is, formation mechanism of discs around Be stars
\citep[e.g.,][]{RiviniusCarciofiMartayan13}. 
Although Be stars are rapidly rotating B-type stars, their rotation velocities are not necessarily 
very close to the
critical ones, above which mass shedding takes place from the equatorial regions of the stars. 
Certain mechanisms are necessary to
accelerate the surface rotation beyond the critical velocities from subcritical ones. 
We suggest that angular momentum transported from the deep interior to the surface by low frequency g-modes excited
by unstable convective modes can be such a help for disc formation around rapidly rotating B stars
\citep[e.g.,][]{AertsMathisRogers19}.  
Efficiency of angular momentum deposition, however, depend on the amplitudes of the excited g-modes, which is another difficult problem to solve.

\bigskip
\noindent
Data Availability: The data underlying this article will be shared on reasonable request to the corresponding author.

\section*{Acknowledgements}

We thank the anonymous referee for his/her very constructive comments on the original manuscript, which 
are very helpful to improve
the presentation of the paper.

\bibliographystyle{mnras}
\bibliography{myref}

\end{document}